\documentclass[twocolumn,pre,floatfix,showkeys]{revtex4}
\usepackage[T1]{fontenc}
\usepackage{amsmath,amsfonts,amssymb}
\usepackage{graphicx}
\usepackage{wrapfig}
\usepackage{placeins}
\usepackage{mathptmx}
\usepackage{color}
\usepackage{array}
\usepackage{dcolumn}

\DeclareMathAlphabet{\mathcal}{OMS}{cmsy}{m}{n}

\renewcommand\Pr[1]{\textbf{P}\left[\:#1\:\right]}
\newcommand\mean[1]{\left\langle{#1}\right\rangle}
\newcommand\var[1]{\mathbf{Var}\left({#1}\right)}

\renewcommand\vec[1]{\mathbf{#1}}
\newcommand\chem[1]{\mathrm{#1}}

\newcommand\konS{k^{+}_{\text{S}}}
\newcommand\konP{k^{+}_{\text{P}}}

\newcommand\koffS{k^{-}_{\text{S}}}
\newcommand\koffP{k^{-}_{\text{P}}}
\newcommand\kcat{k_{\text{cat}}}

\newcommand\msd{\left\langle\|\Vec{p}(t)\|^2\right\rangle}

\newcommand\R{\mathbb{R}}

\newcommand\rarrow[1]{\xrightarrow{#1}}
\newcommand\rrevarrow[2]{\begin{smallmatrix}\displaystyle\xrightarrow{#1} \\
\displaystyle\xleftarrow[#2]{} \end{smallmatrix}}

\newcommand\argmin{\mathop{\text{argmin}}}
\newcommand\tmax{t_{\text{max}}}

\newcommand\unit[1]{\,\mathrm{#1}}
\newcommand\nm{\unit{nm}}
\newcommand\pN{\unit{pN}}

\newcommand\pers{\unit{s^{-1}}}
\renewcommand\sec{\unit{s}}

\newcommand\sci[2]{#1\times10^{#2}}
\newcommand\uvecx{\,\hat{\mathbf{x}}}

\newcommand\velx{\bar{v}_{x}(t)}

\DeclareGraphicsExtensions{.pdf,.eps}
\begin{document}
\pagestyle{empty}
\author{Mark~J.~Olah}
\email{mjo@cs.unm.edu}
\author{Darko Stefanovic}
\email{darko@cs.unm.edu}
\thanks{to whom correspondence should be sent}
\affiliation{Department of Computer Science, University of New Mexico, MSC01 1130, 1 University of New Mexico, Albuquerque, NM 87131-0001}
\title{Superdiffusive transport by multivalent molecular walkers moving
under load}

\begin{abstract}
We introduce a model for translational molecular motors to demonstrate that
a multivalent catalytic walker with flexible, uncoordinated legs can transform
the free energy of surface-bound substrate sites into mechanical work
and undergo biased, superdiffusive motion, even in opposition to
an external load force.
The walker in the model lacks any inherent
orientation of body or track, and its legs
have no chemomechanical coupling other than the passive constraint imposed by
their connection to a common body.
Yet, under appropriate kinetic conditions 
the walker's motion is biased in the direction of unvisited sites,
which allows the walker
to move nearly ballistically away from the origin as long as a local supply of
unmodified substrate sites is available.
The multivalent random walker model is mathematically formulated as
a continuous-time Markov process and is studied numerically.
We use Monte Carlo simulations to generate ensemble estimates of the mean
squared displacement and mean work done for this non-ergodic system.  Our
results show that a residence time bias between visited and unvisited sites
leads to superdiffusive motion over significant
times and distances.
This mechanism can be used to
adapt any enzyme--substrate system with appropriate kinetics
for use as a functional chemical implementation of a molecular motor, without the
need for structural anisotropy or conformationally mediated
chemomechanical coordination.  
\end{abstract}
\maketitle

\section{Introduction}

Motion at the nanoscale is dominated by random, thermally driven
collisions that lead to slow, uncontrollable \emph{diffusive transport}.
Diffusion of large cargo molecules in a crowded cellular environment is
so slow that nature has evolved sophisticated, specialized molecular
machines to transport cargo directionally at superdiffusive rates.
These translational molecular motors, such as kinesin, dynein, and
myosin~\cite{Vale:2000}, walk along \emph{oriented} tracks, consuming chemical
energy in the form of ATP and converting it to mechanical energy that
is used to do work against external load forces~\cite{Svoboda:1993, Vale:1996}.
Molecular motors are essential to a cell's ability to control the
distribution and movement of materials and information within it.

\begin{figure}
\centering
\includegraphics[width=\columnwidth]{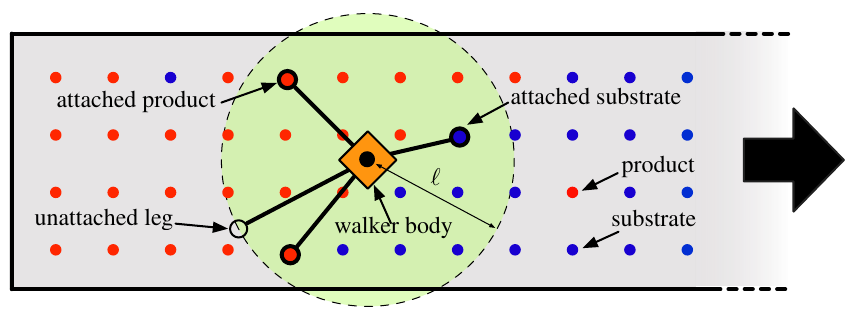}
\caption{(color online) A multivalent random walker (MVRW) is an abstract model of
a multivalent enzyme with a rigid, symmetric body and $k$ identical
enzymatic legs that can reversibly attach to surface-bound chemical sites.
Each leg is flexibly-tethered to the body with maximum extension length $\ell$.
The enzymatic action of a leg can irreversibly transform
a substrate site into a product, changing the subsequent binding kinetics
for the site.  As the legs attach to and detach from sites, the walker moves
over the surface.\label{fig:intro}}
\end{figure}

Synthetic nanoscale systems, like the natural cellular systems that have
inspired their development,
also need mechanisms to maintain non-equilibrium
distribution of materials and information~\cite{Kay:2007}.
Useful synthetic behavior has been demonstrated by combining
natural molecular motors and components in novel ways~\cite{Hess:2010}.
However, each species of natural molecular motor is
highly specialized to its cellular environment, chemical fuel source,
and polymeric track (e.g., microtubules, actin, or DNA).
Synthetic systems with different polymers must either adapt existing
natural motors, or use newly designed compatible synthetic motors.
Kinesin and other natural motors can be mutated to change the kinetics of
their motion~\cite{Thorn:2000}, but not to catalyze arbitrary fuel
substrates, or move over arbitrary tracks without
fundamentally altering their functionality or
efficiency.
Natural molecular motors
rely on non-local conformational changes to couple the binding of fuel
with the kinetics of track binding~\cite{Tomishige:2006}, and use this
conformationally-mediated chemomechanical
coupling to coordinate their processive hand-over-hand
walking gait~\cite{Toprack:2009}.  These mechanisms make natural motors
efficient but also hard to mimic in synthetic systems.

We show that neither oriented tracks, nor rigid walking gaits,
nor chemomechanical coupling, nor coordinated conformational
changes, are necessary for a molecular walker to do
ordered mechanical work as a molecular motor.
We consider the motion of a multivalent walker with a rigid, inert
body and several flexible, enzymatic legs (Fig.~\ref{fig:intro}).
The legs attach to
and enzymatically modify surface-bound chemical sites arrayed as arbitrary
2D patterns and tracks.  The legs are chemically and conformationally uncoupled,
other than by the passive constraint imposed by the connection to a common body.
Yet, under appropriate kinetic conditions, the walker can be made to move
directionally and processively even in opposition to a force.
By modeling and understanding such simple walker systems, we learn which
chemical and mechanical properties of walker-based motors are sufficient
for superdiffusive motion, and which properties are not necessary.

Our multivalent random walker (MVRW) model
is an abstract description of the motion of these walkers.
It takes the form of a continuous time Markov process that describes
the stochastic motion of a walker as it moves over and modifies
surface-bound sites.
The model is designed so that the Markov process can be efficiently simulated,
yet still accurately model the effect of external forces on the physical motion and
chemical action of the walker's body and legs.

Through Monte Carlo simulations we show that walkers with appropriate kinetics
can move superdiffusively in the direction of
unvisited sites over significant times and distances, and can do so while
performing a non-trivial amount of work against an external load.
This effect can be understood from the spontaneous emergence of a
substrate concentration gradient---a boundary
between visited product sites and unvisited substrate sites.
At this boundary, the non-uniform local substrate concentration
combined with the chemical kinetics of the legs and the
constraints that the body places on leg motion lead to a directional bias away from
previously visited sites.
As the legs irreversibly modify substrates to products,
they move this bias-inducing boundary further from the origin.  Hence,
as long as a walker stays proximate to the boundary it moves ballistically
away from the origin.

The MVRW model is inspired by attempts to model the motion of synthetic
DNA-based molecular walkers, called \emph{molecular spiders}~\cite{Stojanovic:2006}.
The spiders are structurally similar to the abstract MVRW walker shown in 
Fig.~\ref{fig:intro}.
Chemically, molecular spiders employ a 
deoxyribozyme/oligonucleotide~\cite{Bonaccio:2004} enzyme/substrate
system, where the deoxyribozyme legs can bind to and modify (cleave) the
oligonucleotide substrates attached
to the surface.  Molecular spiders have been observed to walk processively in
3D environments~\cite{Stojanovic:2006}, and move directionally over  2D nanoscale
tracks~\cite{Lund:2010}.
Abstract models of 1D spider motion were first proposed by
Antal and Krapivsky~\cite{Krapivsky:2007a,Krapivsky:2007b}, who showed that spiders with
rigid nearest-neighbor hopping gaits and idealized kinetics
would experience an effective bias towards unvisited sites.
Subsequent simulations~\cite{Semenov:2011} have shown this \emph{AK spider model}
to exhibit transient superdiffusive behavior as the walkers move between
periods of ballistic and diffusive motion depending on the walker's position
with respect to the boundary between visited and unvisited sites~\cite{Semenov:2011}.
Other work has extended the AK model to study mathematical properties of AK spider
walks in 1D~\cite{Gallesco:2011a,Gallesco:2011b,BenAri:2011}
and 2D~\cite{Krapivsky:2012}; the collective and cooperative behavior
of multi-spider systems in 1D~\cite{Semenov:2011:ms,Semenov:2012:ms,Frey:2013};
and the effect of a load force on the rigid 1D walking gaits of AK-like
spiders under the kinetic rates specific to deoxyribozymes~\cite{Samii:2010,Samii:2011}.

However it remains unclear how the rigid 1D gaits of the AK model and its derivatives
can be implemented at the chemical level.  Indeed, sophisticated mechanisms are
necessary for the coordinated stepping of the natural
cytoskelatal motors like kinesin I and myosin V, which rely on
oriented tracks, conformational switching, and long-range chemomechanical
coordination to achieve directed, hand-over-hand walking gaits~\cite{Vale:2008,Linke:2009}.
Deoxyribozyme-based molecular spiders lack these structural features, and
there is no evidence to show that coordinated hand-over-hand, inchworm,
or nearest-neighbor stepping gaits can be realized directly
by molecular spiders.
The MVRW model removes any assumptions of leg coordination,
and places no constraints on the gaits of the walker legs,
other than the passive constraint imposed by the finite length of legs and
their connection to a common body.  The inherent structural simplicity of the
walkers in the MVRW model makes the model generally applicable
to \emph{any} enzyme/substrate walker system that shares the simple structural motif of
a rigid, inert body with flexibly tethered enzymatic legs like the walker in Fig.~\ref{fig:intro}.
Hence, the MVRW model is not an extension of the Antal-Krapivsky spider models,
nor is it a direct model of deoxyribozyme-based molecular spiders.
Instead it is a general model of uncoordinated, unoriented, enzymatic walkers
that allows us to show that such a simple walker design, using any
enzyme/substrate system with appropriate kinetics, can be made to move
superdiffusively even under the direct
application of a load force to the body, transforming
the chemical free energy of substrate sites into physical work.

The simulation results and analysis of the MVRW model we present show
that mechanisms for designing molecular motors exist
without the need for chemomechanical coupling, conformational coordination,
rigid walking gaits, or inherent orientation of walker and track.
Multivalent random walkers, like natural molecular motors, are Brownian
ratchets~\cite{Peskin:1993} that rectify random molecular motion
into ordered work and directional transport.  Both systems achieve this
rectification by utilizing the chemical free energy of a substrate fuel.
However, the mechanisms by which MVRWs do this are significantly
different from natural motors.
Unlike kinesin I, myosin V, and other natural cytoskelatal motors,
multivalent random walkers move over arbitrarily arranged 2D tracks,
and are able to do so
without inherent orientation or structural asymmetry.
The gaits of a multivalent random walker
are uncoordinated and acyclic, yet the irreversible modification of
surface sites causes an emergent asymmetry in local substrate concentrations
that is able to bias the motion
of walkers, allowing them to move directionally along prescriptive landscapes.
The structural and chemical simplicity of MVRWs is one of their most important
properties as it means that the
conceptual functionality of a molecular
spider is independent of the specific enzyme/substrate system used in their
implementation.
Hence, multivalent random walkers provide a different perspective
for better understanding what structures, properties, and mechanisms are
minimally necessary to turn a molecular walker into a molecular motor.

\section{The Multivalent Random Walker Model}

We model the motion of a multivalent random walker
as a continuous-time, discrete-state Markov process, where
each state transition corresponds to a chemical
reaction between a leg and a surface-bound site.
The model conceptually separates the
timescales of the relatively slow leg-site interactions from the much
faster physical (mechanical) motion of the walker's body and legs.
In this way, only chemical reactions correspond to state transitions
in the Markov process, and the state space remains discrete.
The Markov property is ensured by assuming that the
physical motion of the position of the body and unattached legs
comes to a equilibrium in between successive chemical
reactions.

The MVRW model is 2D and consists of a walker and an environment
of surface-bound sites.  The walker has a rigid, point-like body which
serves as the attachment point for $k$ flexibly tethered legs, each with
maximum length $\ell$.
The environment is described by
a set $\mathbf{S}\subset \R^2$ of immobile chemical sites.
All sites are initially substrates, but they
can be transformed into products by the enzymatic action of a leg.

The state of the Markov process needs only to describe the state of the
\emph{reactive} chemicals in the system, i.e., the species at each chemical site
and the chemical state of each leg.  The state of the surface is defined by the set
$P$ of sites that have been transformed to products;
all other sites in $\mathbf{S}$ are considered to be substrates.
A walker leg is either attached to
a site in $\mathbf{S}$ or is detached.  No two legs may be attached to the same
site.  The state of the walker is succinctly represented by
the set $A\subset \mathbf{S}$ of attached
sites, where $0\leq|A|\leq k$.
Thus, any state $\omega$ can be described compactly as
$\omega=(P, A)$, and we let $\Omega$ be the set of all potential states.

\subsection{Chemical kinetics and state transitions}
\label{sec:kinetics}
In the MVRW model we assume each leg has a single enzymatic site that can bind to
and irreversibly modify a substrate site into product.
The kinetics of an enzymatic leg ($\chem{L}$) binding to substrate ($\chem{S}$) and product
($\chem{P}$) sites can be described by five reaction rates
\begin{equation}
\label{eq:reactions}
\begin{aligned}
\chem{L}+\chem{S} &\rrevarrow{\konS}{\koffS} \chem{LS} \rarrow{\kcat} \chem{L}+\chem{P}+\chem{P^*}\\
\chem{L}+\chem{P} &\rrevarrow{\konP}{\koffP} \chem{LP}
\end{aligned}
\end{equation}
In Eq.~\ref{eq:reactions}, we define the $\kcat$ reaction to encompass both the actual
catalytic cleavage of the $\chem{LS}$ complex together with the subsequent
dissociation of leg $\chem{L}$ from the surface-bound product $\chem{P}$, 
and any other auxiliary product $\chem{P^*}$.
We assume the auxiliary (waste) product $\chem{P^*}$ is not bound to the
surface and its bulk concentration in solution, $[\chem{P^*}]$, is essentially 0.  
Thus, individual rates of binding and unbinding of $\chem{P}^*$ are not important, and the
dissociation reactions can be rolled into the rate $\kcat$.  
The assumption of irreversibility
holds when the Gibbs free energy, $\Delta G$, of the catalysis reaction is
strongly negative, and the rate of the reverse pathway is effectively zero, which
would be the case if $[\chem{P}^*]\approx 0$.

The advantage of this definition of $\kcat$ is that there is a direct correspondence
from reaction rates to Markov process transitions.
The reactions of Eq.~\ref{eq:reactions} each correspond to one of three types
of functional motion for walker legs, association (binding), dissociation (unbinding),
and catalysis.  Each of these actions corresponds directly to a transition in 
the walker Markov process.  In the MVRW model, we assume that rates for the 
unimolecular
dissociation and catalysis reactions for a leg--site complex
are not chemomechanically coupled to the
conformations or positions of the body and other legs.
Thus, as in the unimolecular
stochastic kinetic models of Gillespie~\cite{Gillespie:1976},
each individual $\chem{LS}$ or $\chem{LP}$ complex will dissociate or undergo
catalysis according to the rates $\koffS$, $\koffP$, and $\kcat$, and the time
until that reaction happens will be exponentially distributed according to the
sum of the potential reaction rates from that bound state.

The bimolecular association reactions
are more complicated to model as their propensity depends not only on rates $\konS$ and
$\konP$, but also in the likelihood of the leg being proximate to the chemical
site in order that it may bind.  This likelihood, in turn, depends on the
position of the body and the unattached legs.

\subsection{The body's equilibrium position}
Molecular motors and molecular walkers operate in a regime where they are almost
always at physical (mechanical) equilibrium with their
surroundings~\cite{Bustamante:2000}, and this fact is critical to understanding
how molecular motors operate at the level of discrete chemical transitions~\cite{Lipowsky:2007, Astumian:2010, Lipowsky:2011}.
After each chemical reaction of a leg attaching, detaching, or cleaving, 
the walker's body is subject to high-frequency
thermally-driven constrained diffusion, which quickly brings the walker body to
a physical (mechanical) equilibrium distribution $\Vec{B}$ over 2D positions on the
surface.  This ensures that the Markov property holds for the discrete chemical
states in the MVRW stochastic process because the high-frequency physical motion of the walker quickly
removes any conformational memory of previous chemical states, and the body
distribution $\Vec{B}$ only depends on the current state $\omega=(P,A)$, and not on
any previous states in the Markov process.

\begin{figure}
\centering
\includegraphics[width=\columnwidth]{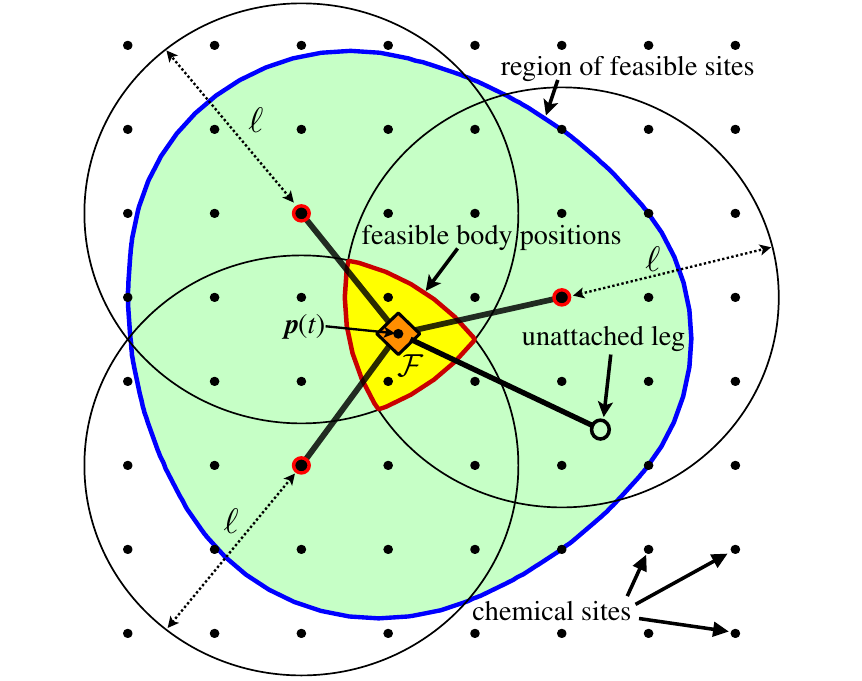}
\caption{(color online) The feasible body positions, $\mathcal{F}$,  as determined by the attached
leg constraints are shown in yellow.  Any site at most distance $\ell$
from $\mathcal{F}$ is a feasible site (green region).
\label{fig:feasible-positions}}
\end{figure}

We assume that the only coupling between the attached legs and the body is that the
body is constrained to stay within distance $\ell$ from each attached site.
Hence, $\Pr{\Vec{B}=\Vec{p}}=0$ for position $\Vec{p}\in\R^2$
if there is any attached leg site $\Vec{s}\in A$
such that $\|\Vec{p} - \Vec{s} \|> \ell$.
We call all values of $\Vec{p}$ that satisfy
$\|\Vec{p}-\Vec{s}\|\leq\ell$ for all $\Vec{s}\in A$ the
\emph{feasible body positions}, $\mathcal{F}$, illustrated in
Fig.~\ref{fig:feasible-positions}.

At equilibrium,
$\Vec{B}$ is a Boltzmann
distribution over the feasible positions $\Vec{p}\in \mathcal{F}$ according to the
energy $U(\Vec{p})$ at each position,
\begin{equation}
\label{eq:body}
\Pr{\Vec{B}=\Vec{p}}=p_{\Vec{B}}(\Vec{p})= \frac{e^{-\beta U(\Vec{p})}}{\int_{\mathcal{F}}
e^{-\beta U(\Vec{p})}d\Vec{p}}.
\end{equation}
In Eq.~\ref{eq:body}, $\beta=1/k_B T$, where $k_B$ is Boltzmann's constant and
$T$ is absolute temperature.  The energy $U(\Vec{p})$ necessarily depends
on the entropic and mechanical properties of the walker legs and their tethers,
the details of which are possible to model in $U(\Vec{p})$, but are dependent on the
actual chemical construction of the walker legs and tethers.
To keep our analysis generally applicable to any flexible tether, we choose
a null hypothesis of no mechanical coupling or internal structure to the legs,
and model the energy $U(\Vec{p})$ as uniform over any feasible
position, but infinite for infeasible positions,
\begin{equation}
\label{eq:U}
U(\Vec{p})=\begin{cases} 0 & \Vec{p}\in\mathcal{F}\\ \infty & \textrm{otherwise} \end{cases}.
\end{equation}

\subsection{Leg--site binding kinetics}

The bimolecular kinetics of leg--site binding is
controlled by two factors, (I) a second-order process by which the leg and
site come into contact, and (II) a first-order process wherein
the leg and site undergo conformational changes to move to a strongly bound
state~\cite{Bustamante:2000}.
We consider the case when the legs are short enough, and the conformational
changes leading to binding are slow enough that factor II is limiting.
In this case, an unattached leg undergoing constrained diffusion has the opportunity
to interact many times with the local feasible sites before it finally binds
strongly enough to be considered attached.
The overall rate of a leg reacting with any feasible site
is then proportional to the number of feasible sites in its proximity.
Equivalently, from the perspective of a feasible site, the probability that
it reacts with the leg is independent of the number of other feasible
sites in the local environment.
For any site $\Vec{s}$ and body position $\Vec{b}$ we define a feasibility function,
\begin{equation}
f_{\Vec{s}}(\Vec{b})=
\begin{cases}
1 & \|\Vec{s}-\Vec{b}\|<\ell \\
0 & \textrm{otherwise}
\end{cases}. 
\end{equation}
Then from position $\Vec{b}$, an unattached leg binds to site $\Vec{s}$,
with species $\pi(\Vec{s})\in\{\chem{S},\chem{P}\}$, with rate
\begin{equation}
\label{eq:rb}
r_{\Vec{b}}(\Vec{s}) = k_{\pi(\Vec{s})}^{+} f_{\Vec{s}}(\Vec{b}).
\end{equation}

Now, we take into account that the body is not at a single position $\Vec{b}$, but
in an equilibrium distribution $\Vec{B}$ over positions, and we integrate
Eq.~\ref{eq:rb} to obtain
\begin{equation}
\label{eq:rB}
r_{\Vec{B}}(\Vec{s}) = k_{\pi(\Vec{s})}^{+}
\int_{\mathcal{F}} p_{\Vec{B}}(\Vec{b}) f_{\Vec{s}}(\Vec{b})\,\, d\Vec{b}.
\end{equation}
Any site with
non-zero rate of attachment is called a
\emph{feasible site}; the region of feasible sites is
shown in Fig.~\ref{fig:feasible-positions}.

\subsection{The effect of force on walkers}
The MVRW model can also capture the effect of forces on the walker body.
This is an advantage of modeling the body's position as a
Boltzmann distribution determined by the energy of the walker at each
feasible position.  Under the effect of a conservative load
force $\Vec{f}$, the energy of position $\Vec{p}$ is
\begin{equation}
U_{\Vec{f}}(\Vec{p})=U(\Vec{p})-\Vec{f}\cdot(\Vec{p}-\Vec{p}_0),
\label{eq:energy}
\end{equation}
where $\Vec{p}_0$ can be any reference point.
The energy $U(\Vec{p})$ is the energy of the body at
position $\vec{p}$ under zero force,
which is defined in Eq.~\ref{eq:U} using our assumption of a
uniform $U(\Vec{p})$.
This choice of $U(\Vec{p})$ represents a worst-case scenario from the
perspective of force production by walkers, as the mechanical structure of the
attached legs is not able to oppose any forces acting on the walker.
Despite this disadvantage,
we will show in Sec.~\ref{sec:results} that walkers can
still move superdiffusively in opposition to a load force applied to the body.

\begin{figure}
\begin{center}
\includegraphics[width=\columnwidth]{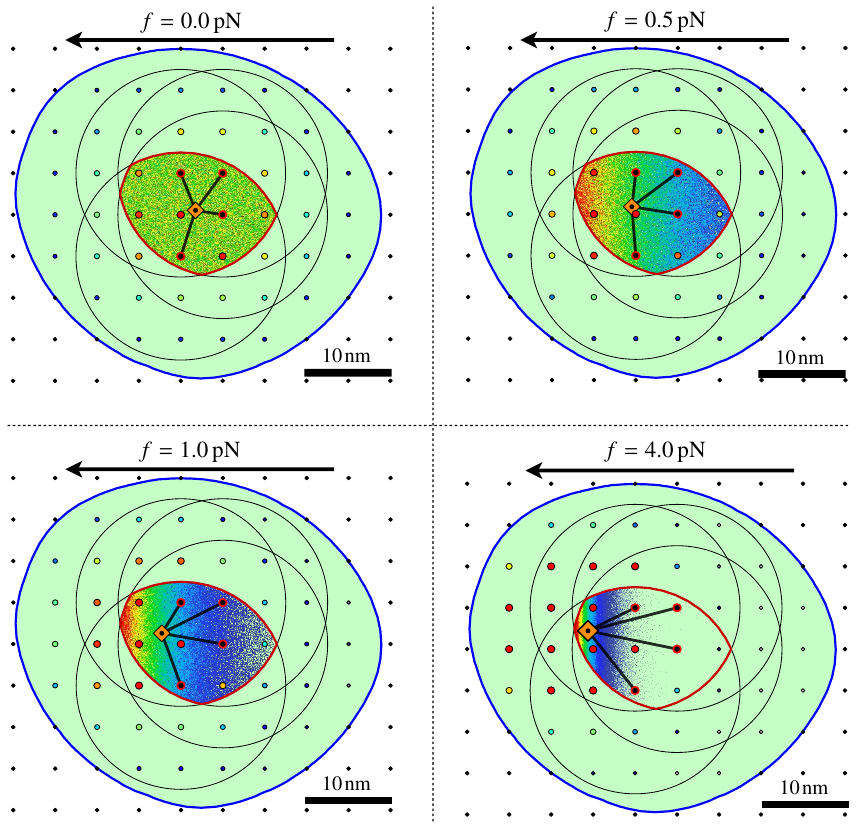}
\caption{(color online) The equilibrium body position probability density
  for a walker under several different load forces.  Warmer colors represent
  increasing probability.  The body is drawn at its mean equilibrium position,
  $\mean{\Vec{B}}$.  The region
  of feasible sites is illustrated in green, but not all sites are equally likely
  for attachment. The color and size of feasible sites
  indicate the effective attachment rate as determined by the body position
  distribution $\Vec{B}$.
\label{fig:body-distribution}
}
\end{center}
\end{figure}

The adjusted energy $U_{\Vec{f}}$ in Eq.~\ref{eq:energy}
gives a new equilibrium distribution
with probability mass shifted in the direction of the applied force.
The effect of force on the body's equilibrium position and on the
attachment propensity for each of the feasible sites is illustrated in
Fig.~\ref{fig:body-distribution}.

\section{Simulation and analysis methods}
The MVRW model is a continuous-time Markov process (CTMP) with
discrete states $\omega=(P,A)\in\Omega$.  Given all relevant parameters,
the walker CTMP defines certain random variables $\{X(t)\}_{t\geq 0}$ over
the state space $\Omega$.
This single Markov process simultaneously describes the fast physical motion of the walker body
and legs under an external load, as well as the slower discrete chemical state changes
of the legs binding and modifying the surface sites.
The MVRW simulation process uses a
kinetic Monte Carlo algorithm~\cite{Bortz:1975,Voter:2007} to sample from the
\emph{non-equilibrium} behavior of the walker's \emph{chemical} motion,
but each chemical step in this MVRW Markov Chain requires
using the Metropolis-Hastings algorithm to sample the \emph{equilibrium}
behavior of the body's \emph{physical} motion.  Hence, by separating the
timescales of the chemical from the physical behavior of the system,
we can take advantage of both equilibrium and non-equilibrium Markov chain Monte
Carlo techniques, using each technique where it is most applicable
to the dynamics of the walkers.

Our kinetic Monte Carlo (KMC) algorithm for the MVRW process produces samples
$x(t)$ for $t\in[0,\tmax]$, such that at each time, $x(t)$ is
a sample of a random variable $X(t)$.
With an ensemble of $n$ samples of the Markov process, we measure and report
various properties of the system state at linearly and logarithmically spaced
time points $t\in[0,\tmax]$.  The simulation algorithm is described
in detail in previous work~\cite{Olah:2011,Olah:2012:dissertation}.

In order to compute the transition rates for the association reactions, we
use the Metropolis-Hastings
algorithm~\cite{Metropolis:1949,Metropolis:1953,Hastings:1970} to sample from
the body's equilibrium
distribution $\Vec{B}$ and use these samples
for Monte Carlo integration of Eq.~\ref{eq:rB}.
Importantly, the Metropolis-Hastings algorithm
is able to sample from $\Vec{B}$ using only the energy from Eq.~\ref{eq:energy},
without having to compute the partition function of the Boltzmann distribution
of Eq.~\ref{eq:body}.

There are computational advantages to modeling the distribution
$\Vec{B}$ at equilibrium.  When the forces on the walker
are conservative, $\Vec{B}$ is translationally invariant,
and depends only on the relative locations
of the attached legs, and not on the whole system state $\omega=(P,A)$.
When the walkers move over \emph{regular} lattices, there are only a finite
number of potential
leg attachment gaits, and their corresponding attachment propensities can be
precomputed, eliminating the need to run Metropolis-Hastings at every KMC iteration.
This makes KMC simulation tractable for long times and large values of $n$.
The details of the simulation of MVRWs on regular lattices can be found in
previous work~\cite{Olah:2012:dissertation}.

\subsection{Random number generation}
In all Monte Carlo methods the fundamental source of stochasticity
derives from a deterministic pseudo-random number generator.
The statistical properties of the pseudo-random number source
are critically important to the correctness of model predictions
~\cite{Ferrenberg:1993}.  To allow parallel computation and
preserve mathematical guarantees of random number generator quality,
we use the leapfrogging
method to generate $n$ parallel random number streams from a single
master stream~\cite{Mertens:2007}.
Hence, only a single random seed is needed to compute
all $n$ KMC traces for each set of model parameters studied.

\subsection{Mean squared displacement and diffusion}
\label{sec:msd-ergodicity}
In single-particle tracking, the stochastic motion of individual molecules
is frequently analyzed in terms of the mean squared displacement (MSD)~\cite{Steinberg:2009}.
The MSD is the variance in the displacement,
$\var{\|\Vec{p}(t)\|}=\mean{\|\Vec{p}(t)\|^2}$.
For any diffusive process (i.e., an unbiased random walk)
the MSD will scale linearly with time. Anomalous diffusion~\cite{Havlin:1987} is characterized by
the MSD scaling as some non-linear power $0\leq\alpha\leq2$,
\begin{equation}
\label{eq:msd}
\msd \propto t^{\alpha},\,\,  \begin{cases} \alpha=0 & \text{stationary}\\
0<\alpha<1 & \text{subdiffusive}\\
\alpha=1 & \text{diffusive}\\
1<\alpha<2 & \text{superdiffusive}\\
\alpha=2 & \text{ballistic or linear}
\end{cases}.
\end{equation}
MSD can either be computed as a temporal average (over different $\delta t$ values
for a single walker trajectory) or an ensemble average (over absolute $t$ for an
ensemble of trajectories from identical walker systems).
Many biological systems
are (or are at least assumed to be) \emph{ergodic} in the sense that the motion
of a walker is independent of its absolute position on the track and does not
depend on its previous motion over a region of that track.
Under the
assumption of ergodicity the temporal and ensemble MSD are equivalent
(assuming sufficient measurement resolution), but when a non-ergodic system is
analyzed, only the ensemble average is meaningful for use in characterizing
anomalous diffusion~\cite{Lubelski:2008, Metzler:2012}.
MVRWs are a nonergodic system because they irreversibly modify the track as
they move over it.  Thus, the motion of the walker depends on its absolute position
on the track and specifically on whether the local sites are products or substrates.
Hence, only the ensemble MSD can be used to study MVRWs.

\section{Results}
\label{sec:results}
\begin{figure}
\centering
    \includegraphics[width=\columnwidth]{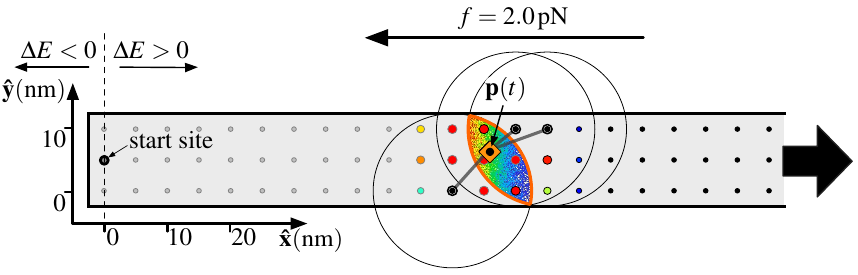}
    \caption{(color online) A snapshot several hundred steps into a MVRW simulation.
The surface track for walkers in this set of simulations
consists of a semi-infinite strip of substrate sites 3-wide.
Shown are the circular constraints imposed by the
attached legs, and the probability density $p_{\Vec{B}}$ for the body's
equilibrium position (as a heat map).
The walker has one unattached leg, and the relative rate at which it would
attach to each site is shown by the size and color of the site.  The boundary of the surface doesn't present any effective constraint on the walker's motion, other than the fact that the lack of sites outside the track prevents the walker body from moving more than a leg length, $\ell$, into those empty regions.\label{fig:forces-illustration}}
\end{figure}

By itself a multivalent random walker is just a rather unsophisticated
multivalent enzyme, but when paired with an appropriately designed nanoscale
track of substrates it becomes a molecular transport device, able to move
superdiffusively even under the influence of an external load force.  Using
KMC simulations we studied the motion of MVRWs moving over a 3-wide
semi-infinite track
of substrates  (Fig.~\ref{fig:forces-illustration}).  The track spacing and other
relevant parameters are summarized in Table~\ref{tab:model-params}. As shown in
Fig.~\ref{fig:forces-illustration}, the walker
starts with a single leg attached to the middle leftmost site at $x$-position 0.
The remaining legs quickly attach, and
the walker begins to move over the surface.
From this initial position, the lack of substrates to the left breaks the symmetry
of the walker's environment, and means the walker can only move in the
$+\uvecx$ direction.  The broken symmetry allows us to apply
a load force to the walker's body in the $-\uvecx$ direction
to oppose the walker's motion, and Eq.~\ref{eq:energy} describes the effect of
that force on the walker body's energy function $U_{\Vec{f}}$.
If the force applied to the walker is $\Vec{f}=(f_x,f_y)$, we let $f_y=0$,
and write $f=-f_x$ as a scalar for the magnitude of the force in the $-\uvecx$
direction.  We limit $f\leq 4.0\pN$ as larger forces
result in insignificant motion under the parameters of Table~\ref{tab:model-params}.
The upper bound of $f=4.0\pN$ is near the maximum force a DNA-based realization
of a MVRW could \emph{a priori} be
expected to move against, as the stall force for kinesin is approximately
$5-8\pN$~\cite{Block:1999}, and the dissociation force for double-stranded DNA
is $<12\pN$~\cite{Essevaz-Roulet:1997}.

\subsection{The role of $\kcat$ in walker kinetics}

\begin{table}
\caption{Model parameters used for simulations.\label{tab:model-params}}
\begin{ruledtabular}
\begin{tabular}{@{\extracolsep{\fill}}lll}
Parameter Description & Symbol  & Value\\
\hline
Number of legs & $k$ & 4\\
Leg length & $\ell$ & $12.5\nm$ \\
Track width & -- & 3 sites\\
Track length & -- & semi-infinite\\
Track site spacing & -- & $5.0\nm\times5.0\nm$\\
Initial set of product sites & $P$ & $\emptyset$ \\
Effective substrate binding rate & $\konS$ & $1.0\times 10^{3}\pers$\\
Effective product binding rate & $\konP$ &   $1.0\times 10^{3}\pers$\\
Substrate dissociation rate & $\koffS$ &$0.0\pers$\\
Product dissociation rate & $\koffP$ & $1.0\pers$\\
Catalysis rate & $\kcat$ & $\leq 1.0\pers$\\
Temperature & $T$ & $300\unit{K}$\\
Force in $-\uvecx$ direction & $f$ & $\leq 4.0\pN$\\
Largest simulated time & $\tmax $ & $\leq\sci{1.0}{7}\unit{s}$\\
\end{tabular}
\end{ruledtabular}
\end{table}

The chemical reactions from Eq.~\ref{eq:reactions} describe the kinetics of
a generic enzyme that can irreversibly transform substrates into products.
From the modeling point of view all five reaction rates, $\konS$, $\konP$, $\koffS$,
$\koffP$, and $\kcat$ are free parameters that can each be varied to determine its
effect on walker motion.  In this work we focus on the special role that $\kcat$
plays in controlling the walker's motion.  Accordingly, we have fixed the values
of the other four kinetic rates as shown in Table~\ref{tab:model-params}, allowing
us to study the effect of varying $\kcat$.  We have fixed $\konS=\konP$ so that there
is no attachment bias between substrates and products.  This choice allows us to focus
on the more subtle kinetic interplay of the remaining rates.  Clearly a walker
with $\konS>\konP$ will be more likely to attach to substrates than products as it
is directly biased in attachment, and we investigate the effect of such kinetics in
Sec.~\ref{sec:param-sensitivity}.
However, we are primarily interested in investigating the minimal
kinetic properties a MVRW must have in order to act as a molecular motor, and so
we assume that $\konS=\konP$.  With these rates equal, an unattached
leg has no ability to differentiate between substrates and products and
will just as rapidly bind to a feasible substrate as to a feasible product.
Yet, even under these conditions a difference in the rates of
$\kcat$ and $\koffP$ can lead to a directional bias.

Consider that, once bound, a leg--product complex unbinds at rate $\koffP$,
and a leg--substrate complex unbinds at rate $\koffS+\kcat$.
We assume that substrate unbinding is
much less probable than substrate catalysis so we let $\koffS=0$ (the relaxation of
this assumption is also considered in Sec.~\ref{sec:param-sensitivity}).
Now, with $\kcat=\koffP$, there is no residence time bias between substrates and
products---the expected duration of a leg--product binding is the same as
that for a leg--substrate binding.
While substrates are still converted into products, the kinetics
of the walker attachment and detachment are identical for both species.  Hence,
a walker with $\kcat=\koffP$ is equivalent to a walker moving over an all-product
surface.  But an all-product surface provides no chemical free energy, and
so an all-product walker system must move diffusively.
Hence, a walker with $\kcat=\koffP$ still releases chemical energy when it
catalyzes the conversion of a substrate, but the symmetrical
kinetics prevent the walker from utilizing that energy.
Thus in subsequent results we have fixed $\koffP=1\pers$ while we vary $\kcat$,
and the case where $\kcat=1\pers=\koffP$ represents the no-energy baseline
motion of walkers.  In contrast, when $\kcat<1\pers=\koffP$ there is a residence
time bias, wherein leg--substrate
bindings are longer in duration than leg--product bindings, as the leg must
wait until the relatively slow catalysis step completes before it can unbind.

The only part of the walker kinetics that takes into
account the chemical free energy released in substrate catalysis
is the assumption of irreversibility in the enzymatic conversion from substrate
to product.  In enzyme kinetics
there is some non-zero rate for the reverse of the catalytic process.
However, if the Gibbs free energy ($\Delta G$)
drop from substrate to product is large enough, the
reverse rate is so small it is for all practical purposes zero,
and is omitted from the walker kinetics in our model (Sec.~\ref{sec:kinetics}).
Thus, we
vary the $\kcat$ parameter to control the residence time bias between visited and
unvisited sites, and at $\kcat=1\pers$ the motion
of the walker is \emph{equivalent} to the no-free-energy case.
We do not directly incorporate $\Delta G$ into the model,
as the kinetic values of $\kcat$ and $\koffP$ are more important
to walker motion than $\Delta G$, and any free energy change
large enough to make the substrate modification effectively
irreversible is sufficient to satisfy the model assumptions.

\subsection{Walkers move superdiffusively in the absence of force}
\begin{figure}
    \centering
    \includegraphics[width=\columnwidth]{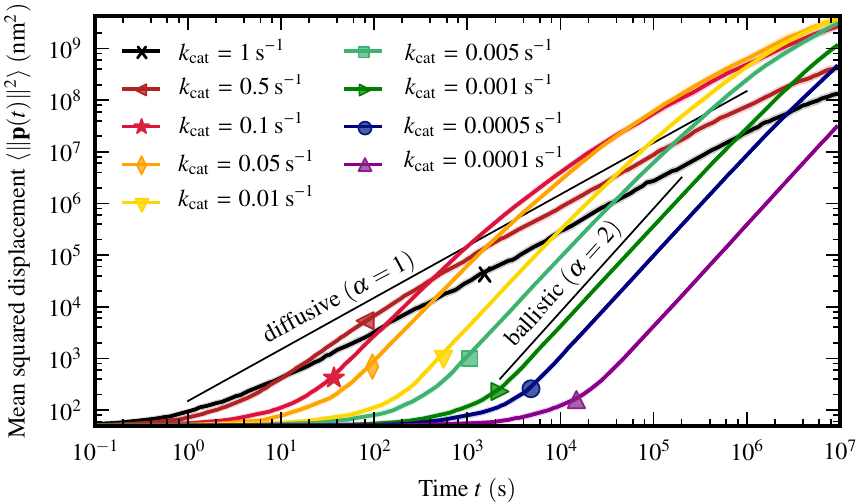}
    \caption{(color online) Simulation estimate of $\msd$ when $f=0$.
    Walkers with $\kcat=1\pers=\koffP$ move
    diffusively.  Those with $\kcat<1\pers$ move superdiffusively, but eventually use up
    their local supply of substrates and become ordinary diffusive.  True transitions to
    diffusion will occur above simulated time $\tmax=10^7\sec$.\label{fig:zero-MSD}}
\end{figure}

Figure~\ref{fig:zero-MSD} shows the ensemble estimates ($n=1000$) for MVRWs
moving in the absence of a load force.
Initially (below the characteristic timescale of $1/\kcat$) the walkers
move subdiffusively.   As expected, the $\kcat=1\pers$
walkers never move faster than diffusion.
However, as $\kcat$ is decreased, walkers initially move more slowly due to
the the slower catalysis kinetics, but once sufficient time has passed, they
move superdiffusively with $\alpha>1$.  The smaller the value of $\kcat$,
the more superdiffusively the walkers move, with $\alpha$ approaching 2 for
the smallest $\kcat$ values.
This superdiffusive behavior
persists over several decades in time, during which
the walkers are moving processively away from the origin in the direction
of unvisited sites.  Because of this outward-directed bias,
the walkers with $\kcat<1\pers=\koffP$ eventually overtake (in MSD) the
$\kcat=1\pers$ walkers given sufficient time.
However, the ability to move superdiffusively depends on
the local availability of the immobile substrate fuel, which is consumed as the
walker moves over the track.  Hence, if a walker moves back
over previously visited
sites, it becomes starved for fuel.  In these energy-devoid regions the walker
can only move diffusively like the $\kcat=1\pers$ walkers, and so superdiffusion
must eventually give way to regular diffusion, even for the smallest values of $\kcat$.

\begin{figure}
    \centering
    \includegraphics[width=\columnwidth]{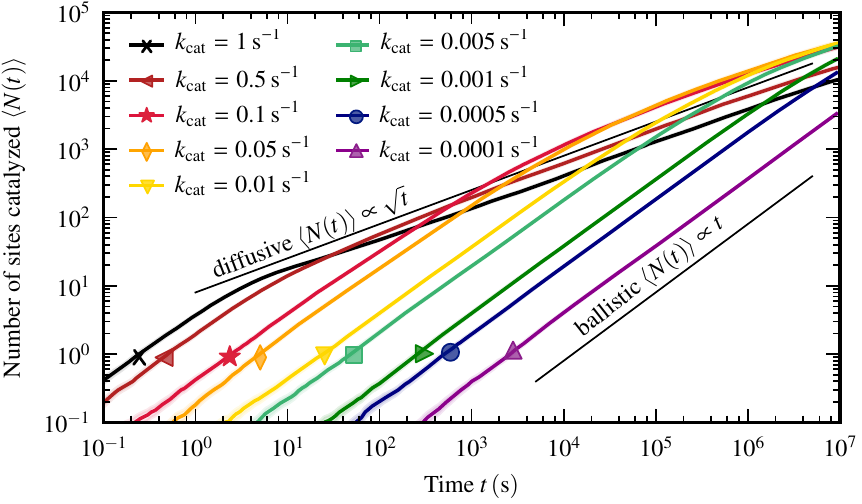}
    \caption{(color online) Simulation estimate of $\mean{N(t)}$, the number of substrates catalyzed to
    products when $f=0$.  Since $\koffS=0$, this is equivalent to the number of
    distinct sites visited at time $t$.  Walkers with $\kcat<1\pers$ catalyze substrates at
    a nearly linear rate over many decades in time. This is necessary to maintain
    a constant supply of chemical energy to sustain superdiffusive motion.\label{fig:zero-NT}}
\end{figure}

Figure~\ref{fig:zero-NT} shows $\mean{N(t)}$, the mean
number of sites catalyzed by time $t$; its rate of change represents the average
availability of substrate fuel.
As long as the number of sites cleaved grows linearly with time,
the walkers are receiving fuel at a constant rate and their motion
is biased in the direction of new sites,
which allows their constant fuel supply to be maintained.
When $\mean{N(t)}$ becomes sub-linear the walkers begin to transition from
superdiffusion to ordinary diffusion.

\subsection{Walkers do work against a load}
\begin{figure*}
    \centering
    \includegraphics[width=\textwidth]{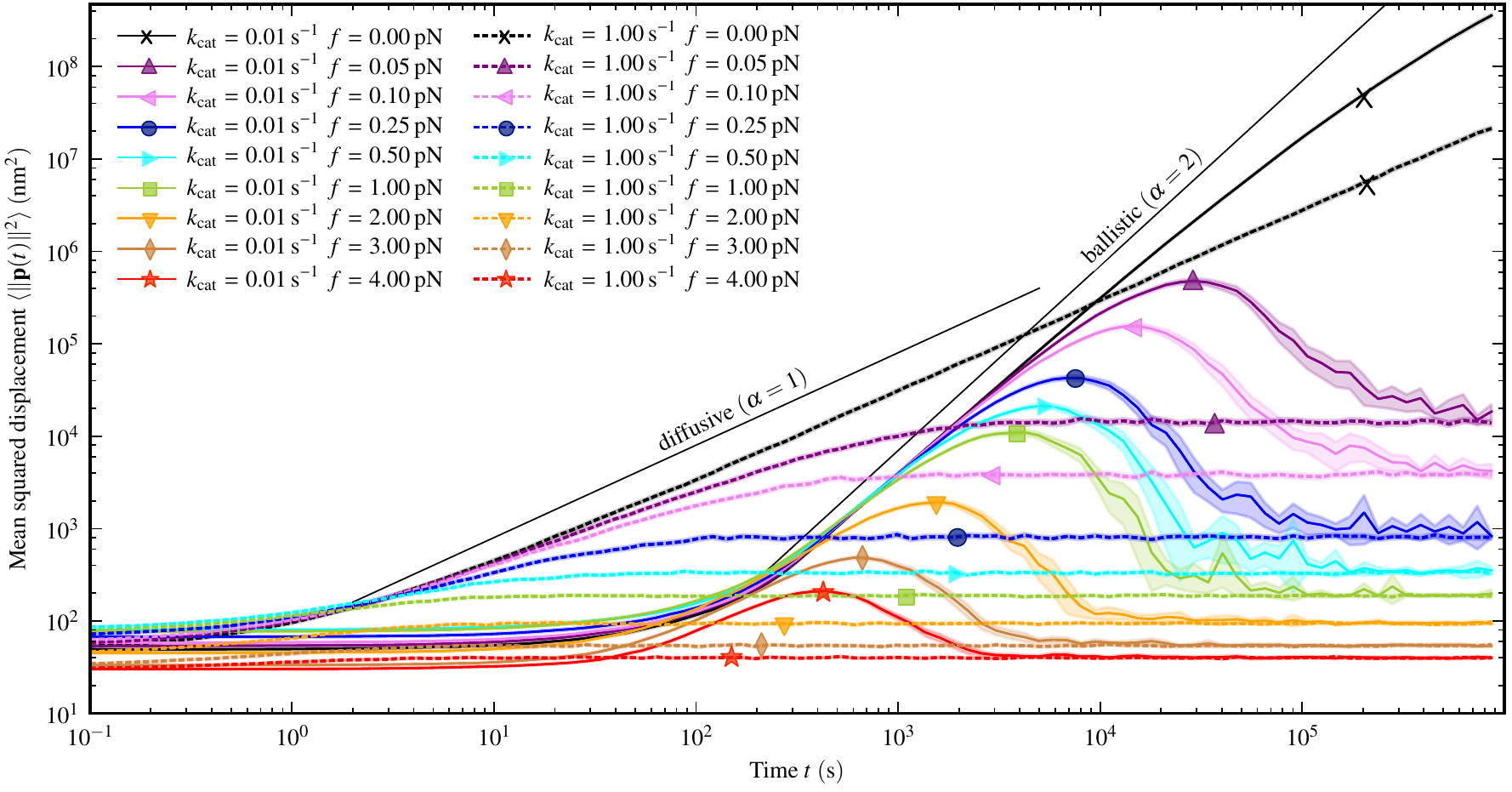}
    \caption{(color online) Simulation estimate ($n=4000$) of $\msd$ and $95\%$
            confidence bounds (shading)
            on a log-log scale. Reference
            lines are shown for ordinary diffusion ($\alpha=1$) and ballistic
            motion ($\alpha=2$).  Walkers with $\kcat<1\pers$ move superdiffusively,
            but when $f>0$, they eventually slow down and return to the same
            equilibrium position as the $\kcat=1\pers$ walkers.\label{fig:forces-MSDLogLog}}
\end{figure*}

To quantify the sensitivity of the walker's superdiffusive motion, we impose a
constant load force $f$ on the walkers in the $-\uvecx$ direction
(Fig.~\ref{fig:forces-illustration}).
Figure~\ref{fig:forces-MSDLogLog} shows ensemble ($n=4000$)
estimates of $\msd$ under a range of forces for $\kcat=1\pers$ and $\kcat=0.01\pers$.
Again, $\kcat=1\pers=\koffP$ (dashed lines) illustrates the no-energy case and,
as shown previously (Fig.~\ref{fig:zero-MSD}),
walkers move diffusively without the influence of force.

\begin{figure*}
    \centering
    \includegraphics[width=\textwidth]{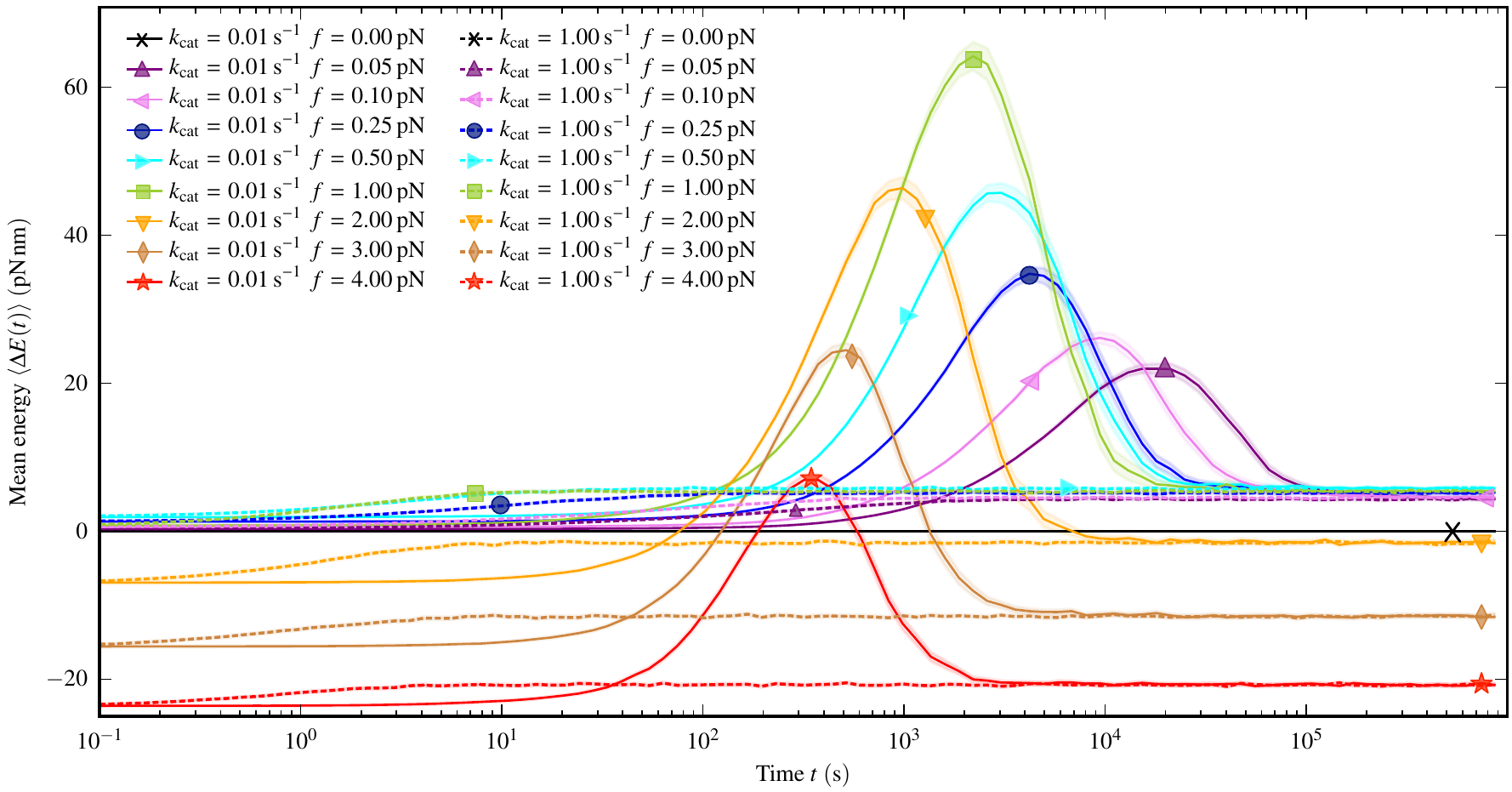}
    \caption{(color online) Simulation estimate ($N=4000$)
    of $\mean{\Delta E(t)}$ and $95\%$ confidence
    bounds (shading) on a log-linear scale.
    Walkers with $f=0$ always have $\Delta E=0$.
    Those with $f>0$ and $\kcat<1\pers$ do significant
    amounts of work, reaching a peak energy before eventually coming to
    an equilibrium.  This equilibrium depends only on $f$ and not on $\kcat$.\label{fig:forces-WorkLog}\vspace{0.5cm}}
\end{figure*}
 
When $f>0$, the random walk over products is biased in the
$-\uvecx$ direction.  The lack of substrates to the left of the origin (Fig.~\ref{fig:forces-illustration}) constrains the walker and the
biased random walk will eventually reach an equilibrium position, after which
the net motion is stationary ($\alpha=0$).  Indeed, this is seen for the $\kcat=1\pers$
walkers, which never move faster than diffusion, and their MSD increases monotonically to
the equilibrium value exactly as if they were undergoing constrained diffusion in a box~\cite{Ritchie:2005}.
In contrast, when $\kcat<1\pers$ we again see nearly ballistic motion for all walkers
except those under the highest load forces $f\geq 2.0\pN$.  Thus, even though
the load force attempts to pull the walker body away from the substrate fuel, the
long residence time for leg--substrate binding allows a few substrate-bound legs
to resist the force and keep the walker in proximity to the substrate sites.
Eventually, as in the $f=0$ case, all walkers regardless of $\kcat$ will
exhaust their local supply of substrates and will find themselves moving over
energy-devoid product sites, which ultimately brings them to the same equilibrium
position as the $\kcat=1\pers$ walkers (for a given $f$).

The change in potential energy of the walkers ($\Delta E$) as they move in
opposition to the load force can be quantified by evaluating the ensemble estimate of
the mean position of the walkers' bodies, $\mean{\Vec{p}(t)}$.
We chose to set $\Delta E=0$ when
$p_x=0$, and then $\Delta E=fp_x>0$ for walkers to the right of the origin
(Fig.~\ref{fig:forces-illustration}).  Figure~\ref{fig:forces-WorkLog} shows
the ensemble estimate of $\mean{\Delta E(t)}$.  As the load force is increased above 0,
the walkers attain progressively higher potential energies, and their peak
energies come earlier, as they need to move less distance to do the same amount
of work.  However, as the forces are increased beyond $f=2\pN$, the walkers
are not able to move very far without being pulled backwards, away from their
substrate fuel, and they achieve only modest values of $\Delta E$.

\subsection{Walker velocity}

Mean walker velocity is another useful measure of walker motility and is commonly used to
characterize the motion of the processive cytoskelatal
motors such as kinesin I and myosin V~\cite{Vale:1997,Howard:1997,Nishiyama:2002}.
Estimation of mean velocity is difficult for multivalent random walkers
because they do not operate in a steady state.
Instead, like other measures of their motility,
mean velocity is time dependent.
Furthermore, the instantaneous velocity between steps has
high variance.

In the experimental setup depicted in Fig.~\ref{fig:forces-illustration},
we are interested in the mean velocity in the $x$ direction, $\mean{v_{x}(t)}$,
as this is the direction in which the force is applied.
Velocity is not a directly observable
quantity of the MVRW model, as walkers move in discrete steps over the state space.
We can directly measure the mean position of the walker $\mean{\Vec{p}(t)}=(\mean{p_x(t)},\mean{p_y(t)})$, which is
defined as the mean location of the body position distribution
$\mean{\Vec{B}}$ (Eq.~\ref{eq:body}).
Due to the variance of random variable $p_x(t)$,
simple finite difference estimations of
$\mean{v_{x}(t_i)}=(\mean{p_x(t_{i+1})}-\mean{p_x(t_i)})/(t_{i+1}-t_i)$ are too noisy
with our sampled data.

In general, computing the derivative of a function known only with noisy measured data is
an ill-posed problem and some sort of regularization procedure must be defined
so that the solution can be uniquely determined~\cite{Vogel:2002}.
The nature of the walker motion implies that
the mean velocity should be a smooth function.
Thus, we follow the methodology of Stickel~\cite{Stickel:2010} in which
the problem is regularized by optimizing for a smooth interpolator
$\hat{p}_x(t)$ that is both a good fit to the data
and that has sufficiently small higher-order derivatives.

\begin{figure}
    \centering
    \includegraphics[width=\columnwidth]{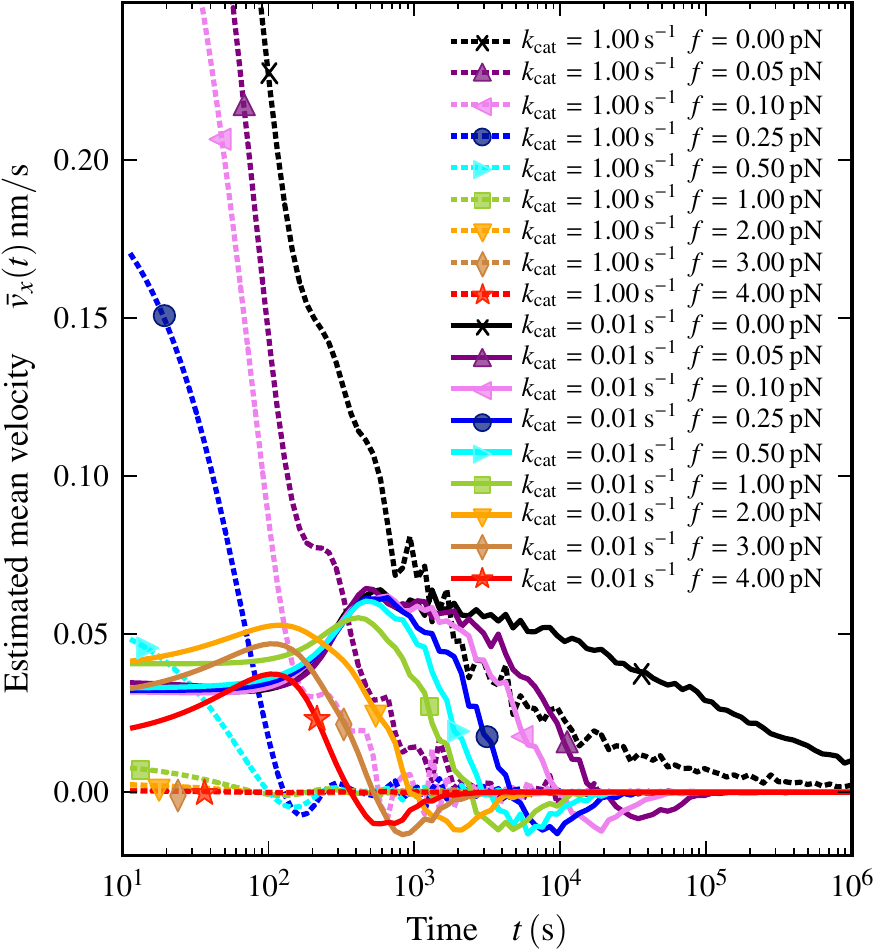}
    \caption{(color online) The regularized finite difference estimate of the
    mean $x$-velocity, $\velx$, for walkers moving under load force $f$ is 
    based on smoothing of the ensemble estimate of $\mean{p_x(t)}$ ($n=4000$).
    Results are shown for times $t\geq t_0=10\sec$,
    as below this time the velocity mainly captures the effect of the
    high-frequency stepping of the walker between adjacent states in the
    state space.  At
    longer times, we can clearly see the net positive velocity away from the origin experienced
    by the $\kcat<1\pers$ walkers even under force.  This net positive motion corresponds
    to the increase in walker energy $\Delta E(t)$ as shown in
    Fig.~\ref{fig:forces-WorkLog}.  The mean velocity approaches 0 as the walkers
    approach the constrained equilibrium imposed by the force.
    \label{fig:vel}}
\end{figure}

Stickel defines a functional $Q$ that ranges over possible smooth
interpolators $\varphi$ on the interval $[t_0, \tmax]$,
\begin{equation}
Q(\varphi) = \int_{t_0}^{\tmax} \left|\varphi(t) - \mean{p_x(t)}\right|^2 \mathrm{d}t + \lambda \int_{t_0}^{\tmax} \left|\varphi^{(d)}(t) \right|^2\mathrm{d}t.
\label{eq:stickelQ}
\end{equation}
In Eq.~\ref{eq:stickelQ} the term $\int_{t_0}^{\tmax} |\varphi(t) - \mean{p_x(t)}|^2 \mathrm{d}t$
measures the $\mathcal{L}^2$-norm of the difference
of the interpolator from the data,
and the term $\int_{t_0}^{\tmax} |\varphi^{(d)}(t) |^2\mathrm{d}t$
measures the $\mathcal{L}^2$-norm of the $d$-th derivative
of $\varphi$.  The smoothed position function, $\hat{p}_x(t)$, is the minimizer
of the functional $Q$,
\begin{equation}
\hat{p}_x= \argmin_{\varphi} Q(\varphi),
\end{equation}
and we can define the smoothed velocity as,
\begin{equation}
\velx = \frac{\mathrm{d}}{\mathrm{d}t} \hat{p}_x(t).
\end{equation}
The weighting parameter $\lambda$ in Eq.~\ref{eq:stickelQ} determines
the relative importance we put
on selecting a $\hat{p}_x(t)$ that minimizes the distance from the
data $\mean{p_x(t)}$, versus
a $\hat{p}_x(t)$ that has small $d$-th order derivative.
As we are looking for the first derivative
of $\hat{p}_x(t)$, we follow the advice of Stickel and optimize for $d=3$,
which is two more than the derivative we require an estimate for.
We found that setting $\lambda=100$ gave an optimal
trade-off between accuracy and smoothness of the resulting derivative,
and these results are shown in Fig.~\ref{fig:vel}.

\subsection{Peak work}
\begin{figure}
    \centering
    \includegraphics[width=\columnwidth]{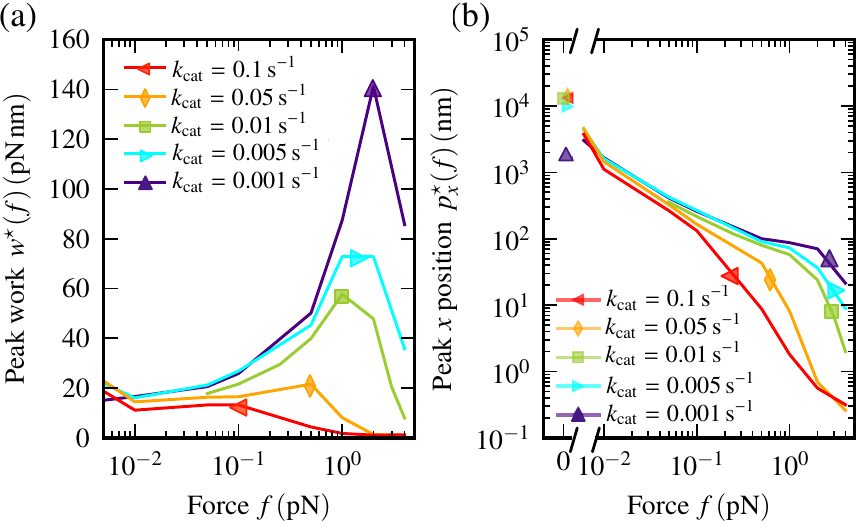}
    \caption{(color online) Simulation estimate of (a) peak work $w^{\star}(f)$ and (b) peak $x$ position
            $p^{\star}_{x}(f)$.  Walkers with $\kcat\in\{1\pers,0.01\pers\}$ are shown with
            an ensemble estimate of the mean using $n=4000$ samples;
            other $\kcat$ are shown using $n=250$ samples.  The peak
            position for $f=0$ is shown as well, which is limited by the
            simulated time $\tmax=1.0 \times 10^6\sec$.  In particular when $f=0$,
            the $\kcat=0.001\pers$
            walkers are still moving superdiffusively at $t=1.0 \times 10^6\sec$,
            but are limited
            by their slower stepping kinetics.  At longer
            times the $\kcat=0.001\pers$ walkers will achieve a peak position greater than
            those achieved by the larger $\kcat$ walkers.\label{fig:forces-peakwork}}
\end{figure}

When $f>0$ all walkers eventually move to an equilibrium position with energy
$\Delta E_{\infty}(f)$.  This value is greater than the initial energy, because the
walkers begin out of equilibrium with only a single leg attached
(Fig.~\ref{fig:forces-illustration}).  The
initial energy of the walker $\Delta E_{0}(f)<0$, because
we measure $\Vec{p}$ as the body's equilibrium position $\mean{\Vec{B}}$,
which under any non-zero force will have $p_x<0$ at the initial walker attachment
location.
However, the kinetics of $\konP\gg\koffP$ lead to an equilibrium
where legs are almost always attached to a site, and because all sites are to the
right of the origin, the equilibrium position $\Delta E_{\infty}(f)$
will also necessarily be greater than $\Delta E_{0}(f)$.
Thus, to characterize the amount of useful work that a walker can do we take
into account the equilibrium energy specific to each force.  We define
the peak work for force $f$ as
\begin{equation}
w^{\star}(f)=\max_{t\in[0,\tmax]} \mean {\Delta E (t;f)} - \Delta E_{\infty}(f).
\end{equation}
We estimate $\Delta E_{\infty}(f)$ as $\mean{\Delta E(\tmax;f)}$ for the
$\kcat=1$ walker.
Figure~\ref{fig:forces-peakwork}a shows $w^{\star}$ as force and $\kcat$ are
varied.  The $\kcat=1\pers$ walkers never have $w^{\star}>0$, but
the walkers with $\kcat<1\pers$ can do significant work under moderate forces.

Figure~\ref{fig:forces-peakwork}b, shows the values for the peak $x$-position,
\begin{equation}
p_{x}^{\star}(f)=\max_{t\in[0,\tmax]} \mean {p_x(t;f)} - p_{x}^{\infty}(f).
\end{equation}
Again we estimate the equilibrium $x$-position,
$ p_{x}^{\infty}(f)$ using $p_{x}(\tmax;f)$ as measured for the $\kcat=1\pers$ walkers.
These measurements show that the walkers
move significantly farther under small loads, although they do nearly the same work.

\section{Discussion}
Multivalent random walkers are able to do work because they
act as Brownian ratchets.  The physical motion of the walker is the result of
random thermally driven molecular motions that are rectified by the constraints
imposed by attached legs.  Without any structural or conformational coupling,
the independently operating legs are constrained only by their passive
connection to a common body.  The gaits with which MVRWs move are uncoordinated,
unoriented, and acyclic, yet they
can be designed to move nearly ballistically along tracks laid out in 2D.

From a thermodynamical perspective,
the walkers are modeled as a closed system, where the only
energy available to the walker is present in the uncleaved substrate sites.
Any closed system will eventually approach a thermodynamic equilibrium after
which no useful work can be accomplished.  Indeed, we see this effect for the
walkers under load force $f$ shown in
Figs.~\ref{fig:forces-MSDLogLog}~and~\ref{fig:forces-WorkLog}, where the walkers
with $\kcat<\koffP$ are able to move superdiffusively over significant distances
and hence do work as they move in opposition to the load force, but they do so
only while they still have energy available to bias their motion.
Eventually, these walkers
move to the same equilibrium distribution as the $\kcat=\koffP$ walkers which
correspond to the no-energy case.

The key concept in the MVRW model of molecular walker motion is that 
energy is a local resource, and the walker depletes the local energy supply as
it moves over a region and catalyzes the conversion of the substrate sites to
products.  This concept makes the MVRW system \emph{non-ergodic}
in the sense that the behavior
of the walker depends on the local distribution of substrates and products,
and this
distribution in turn depends on the past motion of the walker over that region of
the track.  Most natural motors can be described as ergodic, as their fuel source
(normally ATP) is present in solution and the track they move over
is unmodified by their previous actions.
Because MVRWs are non-ergodic they do not operate in a steady state, and
unlike models of natural motors~\cite{Astumian:2010, Lipowsky:2007} there is no
way to quantify the motion of MVRWs by studying a particular set of
cycles of states in their state space.
As mentioned in Sec.~\ref{sec:msd-ergodicity}, the non-ergodicity requires us
to use the ensemble formulation of MSD.  Additionally, most other
random variables that describe MVRW motion ($\mean{N(t)}$, $\velx$,
$\mean{\Delta E(t)}$, etc.) are time-dependent.
Hence, there is no single value
of velocity, or single stall force that can be calculated for MVRW-like systems,
as is normally done for the ergodic motion of natural molecular motors~\cite{Vale:1997}.

\subsection{Mechanism of superdiffusive motion}
\begin{figure}
    \centering
    \includegraphics[width=\columnwidth]{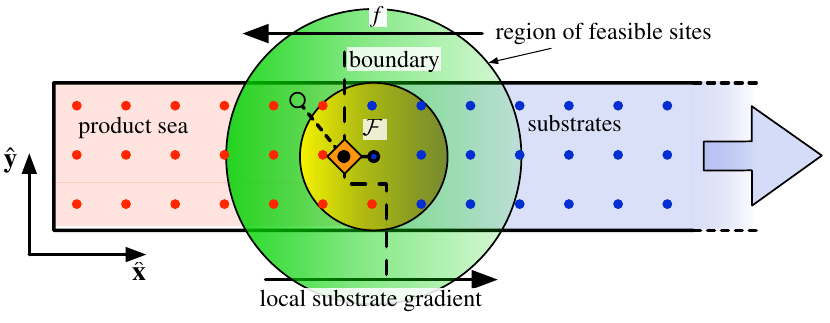}
    \caption{(color online) The irreversible catalysis of substrates to products leads to the
    emergence of a spatial asymmetry in substrate concentration at the \emph{boundary}
    between the contiguous $\emph{product sea}$ and the contiguous region of
    unvisited substrates.
    A walker with $\kcat<1\pers$ has a residence time bias where leg--substrate
    binding durations are much longer than for leg--product bindings.
    Thus, over time, legs are more likely to be attached to local substrates
    than local products, not because they seek out substrates, but because
    legs attached to products quickly detach.
    Hence, walkers are effectively driven in the direction of
    greatest local substrate density, and near the boundary this
    is always in the $+\uvecx$ direction.  The irreversibility of substrate
    catalysis means the boundary itself also moves in the $+\uvecx$ direction,
    causing walkers near the boundary to move ballistically away from the origin.
    \label{fig:Boundary}}
\end{figure}

The superdiffusive motion of walkers and its eventual decay to diffusion ($f=0$)
or stationary equilibrium ($f>0$) can be understood by noting that the only
source of energy available to the walkers is present in the substrate molecules,
which are a locally-limited, immobile resource.

After the walker starts
moving and catalyzing sites, a contiguous
region of product sites we call the \emph{product
sea} begins to form (Fig.~\ref{fig:Boundary}).
At the \emph{boundary} between the product sea and unvisited substrates,
the local substrate concentration gradient is in the $+\uvecx$
direction, due to the broken symmetry introduced by the semi-infinite
surface configuration studied (Fig.~\ref{fig:forces-illustration}).
The emergence of spatial asymmetry in concentration makes it possible for
an unoriented, symmetric walker to develop a directional bias.
At the boundary, a MVRW with $\kcat<1\pers$ is biased in the $+\uvecx$ direction
not because the legs are more likely to attach to substrates ($\konS=\konP$),
but because when they do attach to a substrate, they stay bound longer---there 
is an effective \emph{residence time bias}.

A walker with  $\kcat<1\pers$ is only directionally biased when near the
boundary, in which case its legs irreversibly
catalyze attached substrates to products, moving the boundary in the
$+\uvecx$ direction as well.  Thus, as long as a walker remains
near the boundary, it is biased in the $+\uvecx$ direction, and it moves the
bias-inducing substrate concentration asymmetry along with it, which leads
to persistent motion directed away from the origin.

This mechanism of residence time bias leading to a directional bias was identified
by Antal and Krapivsky in their abstract 1D molecular spider model~\cite{Krapivsky:2007b}.
Later it was shown to lead to significant superdiffusive behavior of the Antal-Krapivsky (AK)
walkers in 1D without force~\cite{Semenov:2011}.
The simple state space of the AK walker models allows it to be shown analytically
that the motion of AK walkers is ballistic in the direction of substrates while they remain proximate to the boundary~\cite{Semenov:2011}.
The 2D geometry of the MVRW model makes the mathematical
description of boundary between substrate and products more complex,
but with the simulation results
in Figs.~\ref{fig:zero-MSD}~and~\ref{fig:forces-MSDLogLog}, we find the motion
in 2D (at the ensemble level) is nearly ballistic even when opposing small forces.
This implies that individual walkers near the boundary must on average
also be moving nearly ballistically, even under the effect of a constant load force.

\subsection{The boundary and diffusive metastates}
\begin{figure}
    \centering
    \includegraphics[width=\columnwidth]{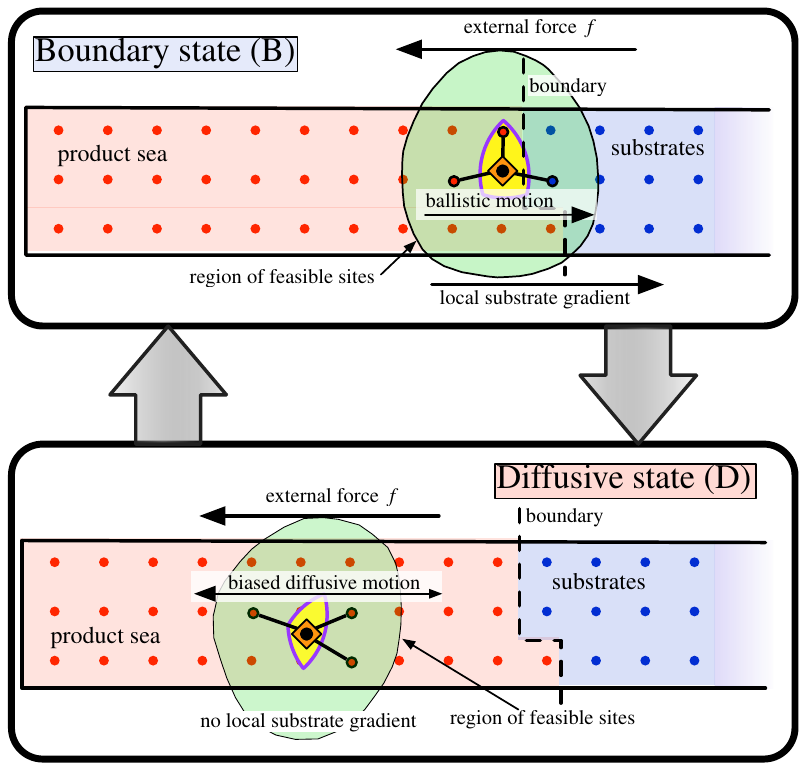}
    \caption{(color online) The walker moves between boundary ($B$) and
    diffusive ($D$) metastates.
    The walker moves ballistically in the direction of local substrate gradient
    when in the $B$ state, but moves diffusively over previously visited sites in
    the $D$ state.  The walker initially spends most of its time in the $B$ state,
    consuming substrate fuel, however as the product
    sea grows, the time to exit the $D$ state increases, leading to asymptotically
    diffusive motion in the absence of force and equilibrium stationary motion in the
    presence of force. \label{fig:BDStates}}
\end{figure}

The emergence of the boundary between the product sea and the unvisited substrates
causes the walker to move superdiffusively, but eventually all walkers either move
diffusively ($f=0$) or move to a stationary equilibrium distribution ($f>0$).
In analogy to our analysis of the AK spider model~\cite{Semenov:2011},
this behavior can be understood by decomposing
the Markov process into two metastates: a boundary state ($B$) wherein the
walker is attached to substrates near the boundary, and a
diffusive state ($D$) wherein the walker moves over the energy-devoid product sea
(Fig.~\ref{fig:BDStates}).

\begin{figure}
    \centering
    \includegraphics[width=\columnwidth]{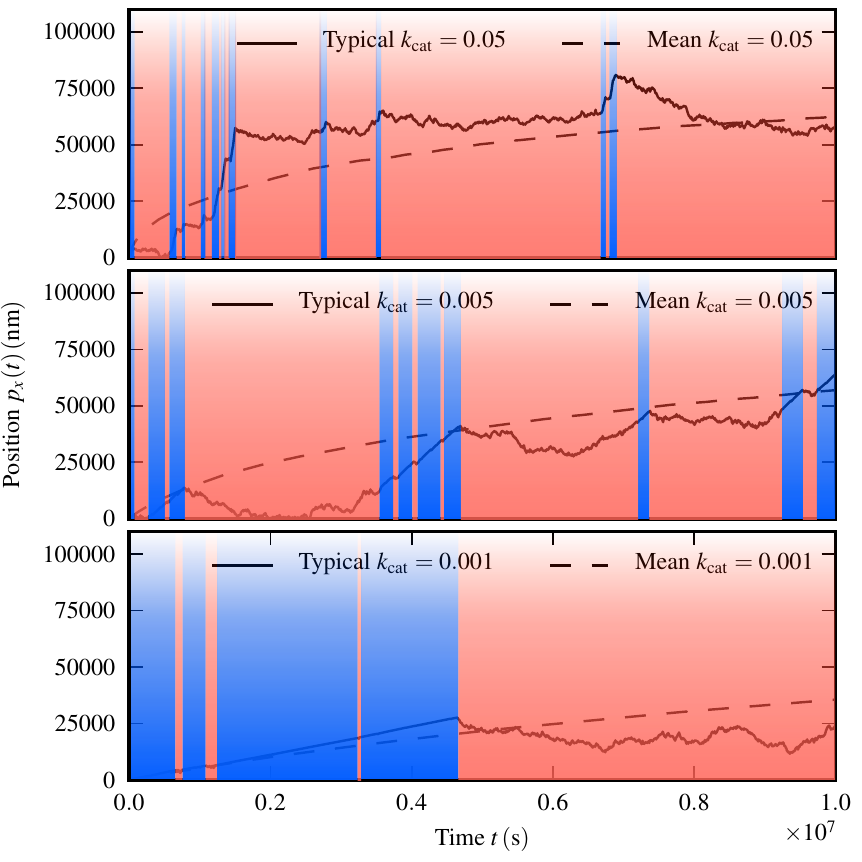}
    \caption{(color online) Typical traces of $p_x(t)$ for a MVRW with $f=0$
    for three $\kcat$ values.
    The traces are shaded blue when the walker is in the $B$ metastate, and red
    when it is in the $D$ metastate.  Walkers with smaller $\kcat$ have longer
    $B$ periods, but smaller velocity.  The duration of $B$-periods
    is independent of time, but the duration of $D$-periods grows with
    the size of the product sea, and consequently increases over time.
    Thus, at short times the walker is more likely to be in the $B$ state, but
    at longer times is more likely to be in the $D$ state.
    \label{fig:BDtypical}}
\end{figure}

When the walker is in the $B$ state it moves ballistically in the $+\uvecx$
direction, but when it is in the $D$ state it has no directional orientation,
and it moves by ordinary unbiased diffusion for $f=0$, or by
$-\uvecx$-biased diffusion when $f>0$.  Figure~\ref{fig:BDtypical}
shows three typical traces of the position of individual walkers under no force,
where $B$ and $D$ periods have been shaded to show the alternation between
states and the distinction between the ballistic and diffusive motion.

The probability of a walker leaving the $B$ state by moving sufficiently far
in the $-\uvecx$ direction is independent of the absolute position of the
boundary.  Thus, the $B$ metastate is Markovian since the transition rate to the
$D$ metastate is independent of how long the walker has been moving or the size
of the product sea.
As $\kcat$ is decreased, the duration of leg--substrate bindings relative to
leg--product bindings increases
and the walker is less likely to simultaneously detach from all boundary
substrates and leave the $B$ state.
Thus, lower $\kcat$ values result
in more persistent ballistic motion over longer durations,
but at smaller velocities (Fig.~\ref{fig:BDtypical}).

In contrast the $D$ metastate is non-Markovian.  The duration of a $D$ period
depends on the size of the product sea, and hence this duration grows as the
walker catalyzes sites.
In the case where $f=0$, the time is quadratically dependent
on the size of the product sea, but when $f>0$ this dependence becomes
exponential, and for sufficient forces and sufficiently sized product seas,
the probability of returning to the boundary once departed a significant
distance becomes effectively 0.
Hence, the duration of $B$-periods is constant in time,
but the duration of $D$-periods grows.
Eventually walkers spend nearly all their time moving over products in the
$D$ state, and so approach the same equilibrium distribution as the $\kcat=1\pers$
walkers which represent the case where no energy is available to the walker.
This eventual drift toward equilibrium can be seen in Figs.~\ref{fig:forces-MSDLogLog}~and~\ref{fig:forces-WorkLog}.

\subsection{Dissociation}
There is a non-zero probability for a walker to detach from the track if
$k-1$ legs are simultaneously in the detached state, and the next action
chosen is for the remaining leg to detach.
A walker with $k$ detached legs is free to diffuse in solution,
and cannot be ascribed a well-defined position with
a discrete state Markov process.
Hence, dissociation poses mathematical difficulties
for analyzing a non-ergodic motive process and comparing
it with other mathematical models of anomalous diffusion.
Ergodic models of natural motors like kinesin I
can simultaneously analyze motion and dissociation,
because the transport characteristics and dissociation probabilities
can be understood independently
by studying a single motor cycle~\cite{Astumian:2010, Lipowsky:2008}.
MVRWs, being non-ergodic, have transport and dissociation probabilities
that depend on the current state of the local chemical sites, and cannot be
analyzed with similar techniques.

One approach to dealing with dissociation
in non-ergodic walker models is to have a single absorbing dissociated state
to which all walkers will eventually go and never return.  This state is then
the single equilibrium state of the system, and analysis is done on the remaining
walkers.  However, analyzing MSD becomes challenging
because at any $t>0$ there is necessarily some non-zero proportion of walkers
in the dissociated state.  Ensemble MSD is no longer well-defined,
as we cannot ascribe a position to dissociated walkers.
Instead of this approach, we implement a \emph{hopping rule},
whereby a walker with $k-1$ legs
whose next KMC chosen transition is to detach its one remaining leg is prevented
from diffusing away from its dissociation location.
It is temporarily held in place until a leg attaches to
a local feasible site.  The net effect is a hop from one site to another
and is implemented as a single KMC step, so that the position of the walker
is always well defined.

For any finite $\konS$ and $\konP$ rates,
it is possible for walkers to temporarily dissociate via a hopping event.
In practice, however,
when the walker has sufficiently many legs, the on-rates are sufficiently fast,
the legs are long, and the substrates are densely spaced, the probability of
dissociation is low.  Over the course of the simulations shown in
Figs.~\ref{fig:forces-MSDLogLog}~and~\ref{fig:forces-WorkLog}, only four out of 56000
walkers with $f<3.0\pN$ and 100 out of 16000 walkers
with $f\geq 3.0\pN$  experienced any hopping event.

\subsection{Effect of variation of number of legs and leg length}
As summarized in Table~\ref{tab:model-params}, our results
focus on four-legged walkers with leg
length $\ell=12.5\nm$, which is 2.5 times the $5.0\nm$ substrate spacing
distance.  Both the leg length and the number of legs
can be freely varied.  However,
there are sensible ranges for these parameters,
outside of which the motion of the walkers is not as processive, or is exceedingly
slow.   To be efficient molecular transport devices, walkers
must simultaneously avoid dissociation, resist the effect of forces,
and remain attached to substrates near the boundary.

\begin{figure}
    \centering
    \includegraphics[width=\columnwidth]{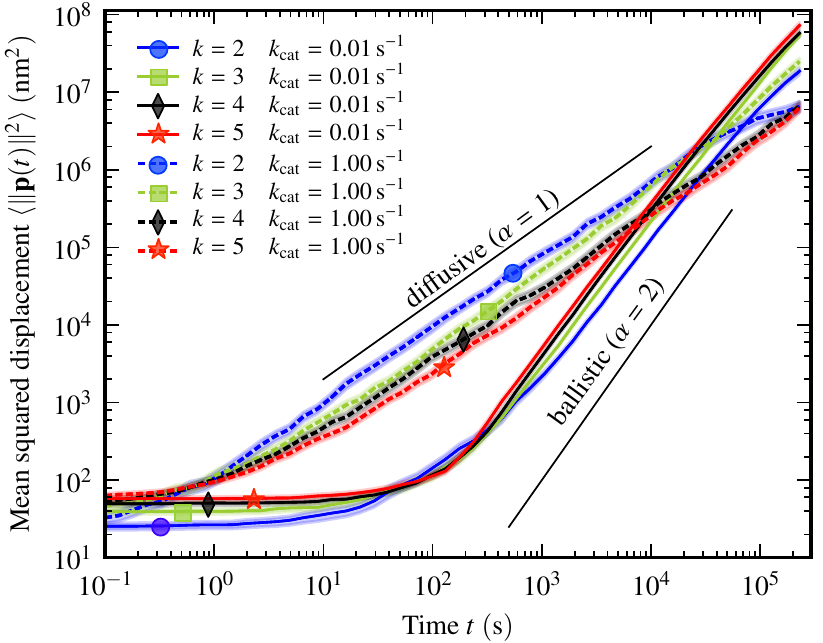}
    \caption{(color online) Simulation estimates $(n=400)$ showing the effect of the number
    of walker legs ($k$) on walker motion when $f=0$.
    The MSD is shown with 95\% confidence intervals in shading.  Walkers with
    more legs move with smaller diffusion constant when $\kcat=1\pers=\koffP$
    and there is no residence time
    bias.  However, when $\kcat=0.01\pers$, the walkers with more legs experience a stronger
    directional bias towards the local substrate concentration gradient and hence move
    superdiffusively over longer times and distances.  Of the configurations studied,
    walkers with $k=5$ legs
    and $\kcat=0.01\pers$ eventually achieve the greatest mean squared displacement.
    The black lines
    show the case $k=4$, corresponding to walkers in Fig.~\ref{fig:zero-MSD}.
    \label{fig:13}}
\end{figure}

First, consider the number of legs, which
is varied in the range $2\leq k\leq 5$
in Fig.~\ref{fig:13}.
For the residence time bias to lead to a directional bias,
we require $k\geq2$~\cite{Semenov:2011}.  With few legs ($k=2$), walkers are
more likely to have all of their legs detached simultaneously and undergo a hopping
step.  As the number of legs is increased this probability drops exponentially,
as each leg's probability of detachment is approximately independent.
Walkers with more legs also tend to
move more superdiffusively and processively when in the $B$ state, as they
have a higher probability that at least one leg remains attached to a substrate at
the boundary.
However, walkers with many legs have a significantly smaller diffusion
constant.
Hence, $k=4$ was chosen as a reasonable compromise value that prevents dissociation,
maintains a strong tendency to remain on the boundary, and moves appreciably fast.

The leg length $\ell$ must be considered in relation to
the substrate spacing.
Together these parameters determine the
average number of feasible sites available for attachment.
Ideally, the substrate spacing would be made as small as possible,
but it is constrained by limits on the sizes
of molecules and how closely substrates can be arrayed on a surface.
We chose $5.0\nm$ as a reasonable lower limit on this spacing, as it approximates the
density of DNA substrates arrayed on a DNA origami~\cite{Rothemund:2006}
surface, as employed in molecular spider experiments~\cite{Lund:2010}.

\begin{figure}
    \centering
    \includegraphics[width=\columnwidth]{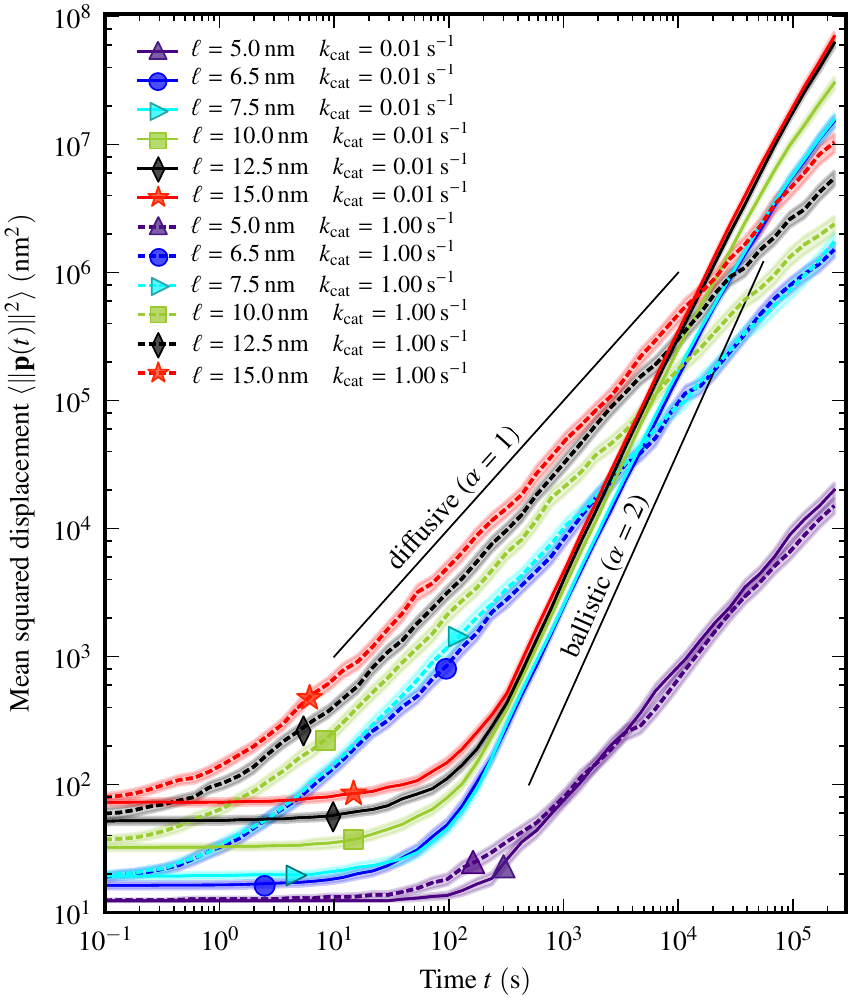}
    \caption{(color online) Simulation results $(n=400)$ showing the effect (at $f=0$) of
    varying the leg length $5.0\,\nm\leq \ell\leq 15.0\,\nm$, while the number
    of legs is fixed
    at $k=4$ and the substrate spacing is fixed at $5.0\,\nm \times 5.0\,\nm$.
    The effect of $\ell$ on MSD is shown with shading indicating the 95\% confidence
    interval for the mean.  The effect of changing the leg length is essentially manifested
    as a change in the diffusion constant, but not in the qualitative characteristics
    of the superdiffusive motion for the $\kcat=0.01\pers$ walkers.  The exception is
    for the very short leg length $\ell=5.0\,\nm$, where the average number of feasible
    sites becomes so small that walkers lose their superdiffusive transport
    behavior\label{fig:14}}
\end{figure}

Figure~\ref{fig:14} shows the effect of varying the leg length for
4-legged walkers while keeping the substrate spacing constant at $5.0\nm$.
We find that if legs are too short ($\ell \leq 5.0\,\nm$), the
number of feasible sites is too small to maintain a superdiffusive
effect.  For leg lengths $\ell\geq 7.5\,\nm$, which is 1.5 times the substrate
spacing, there is little qualitative difference in the
walker motion, although longer legs do lead to a faster diffusion constant in the
absence of force.
Under load, however, leg length and
substrate spacing should both be minimized to maximize the peak work and displacement
of walkers.  Longer legs allow
a larger feasible region $\mathcal{F}$, leading to a larger bias
in $\Vec{B}$ under any non-zero load.  This in turn makes it more likely for
long-legged walkers to move backwards.
We found that a leg length of approximately 2.5 times
the substrate spacing provides a good balance between
dissociation and processivity, although a full analysis of this relationship
is reserved for future study.

\subsection{Sensitivity to kinetic parameters}
\label{sec:param-sensitivity}
The MVRW model has too many independently variable kinetic parameters to simultaneously
examine the effect of each of them on walker motion characteristics.
We have chosen representative kinetics summarized in Table~\ref{tab:model-params}
to act as a reference point.  Clearly, any chemical implementation of the
multivalent random walker model (e.g., molecular spiders) will have potentially
very different (and likely much faster) rates than those we have chosen.  However,
it is not our purpose to model a specific chemical implementation.
Instead, we show that the qualitative characteristics of superdiffusive walker motion
persist over a wide range of kinetic values, as long as the residence time bias
between visited and unvisited sites is present, leading to an effective motive bias
in the direction of the local substrate concentration gradient.

\begin{figure}
    \centering
    \includegraphics[width=\columnwidth]{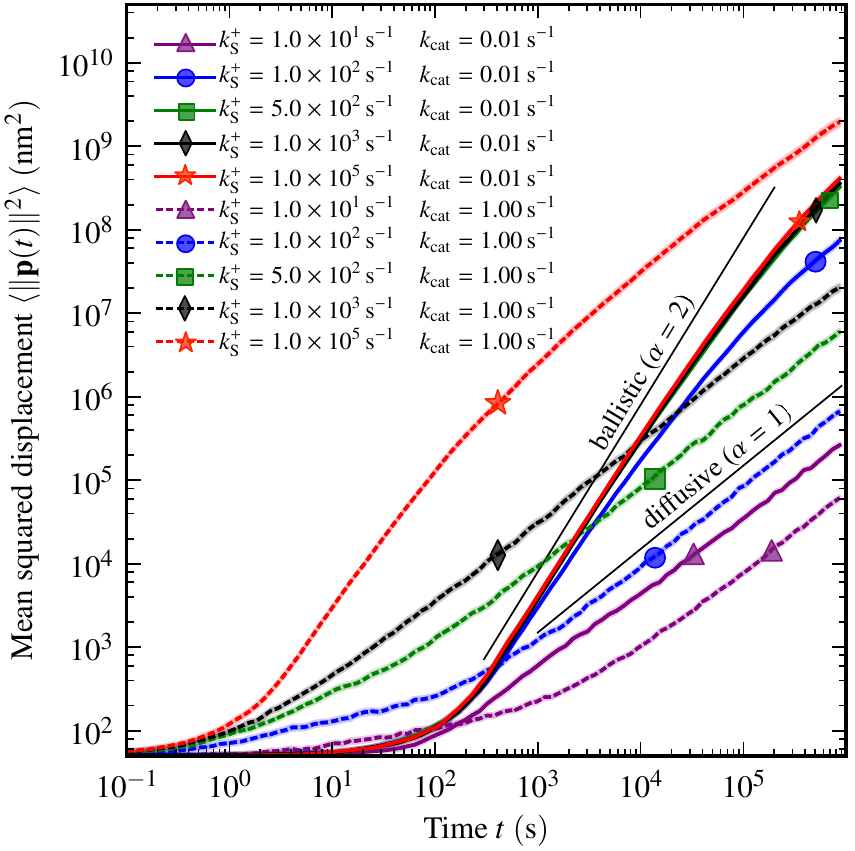}
    \caption{(color online) Simulation results $(n=1000)$ showing the effect (at $f=0$)
    of varying $\konS$ on
    the MSD of walkers while $\konP$ is fixed at $1.0\times 10^3\pers$.
    The black lines represent the case where $\konP=\konS=1.0\times 10^3\pers$, which
    is used in all other simulations results.  When $\konP=\konS$, there is no
    attachment preference for a substrate over a product.  An unattached leg will just
    as rapidly bind to a feasible substrate as to a feasible product.  However, as
    we take $\konS<\konP$, the walkers have an attachment bias to products, which
    should be expected to reduce the time spent in the boundary ($B$) metastate, and
    therefore lead to less pronounced superdiffusive behavior.  These results show that
    the MSD is robust to moderate changes in the on-rates, and
    even for $\konS=\konP/10$, there is an appreciable superdiffusive effect when
    $\kcat=0.01\pers$.  However taking $\konS=\konP/100$ overwhelms the residence time bias
    of the walkers in the $B$ state and prevents any superdiffusive motion.\label{fig:15}}
\end{figure}

In Fig.~\ref{fig:15} we show that the superdiffusive behavior as quantified
by MSD persists over an order of magnitude in variation of the $\konS$ rate.
Indeed, even if the walker is biased 10:1 in attachment preference to products
over substrates, the residence time bias of 100:1 of substrate to product binding
duration is still
sufficient to achieve a superdiffusive scaling of MSD over several decades in
time.  This robustness even to large changes in attachment
rates allows us to be confident
that superdiffusive behavior is a pervasive feature of multivalent random walker
systems and is not critically dependent on our particular choice of attachment
rates.

\begin{figure*}
    \centering
    \includegraphics[width=\textwidth]{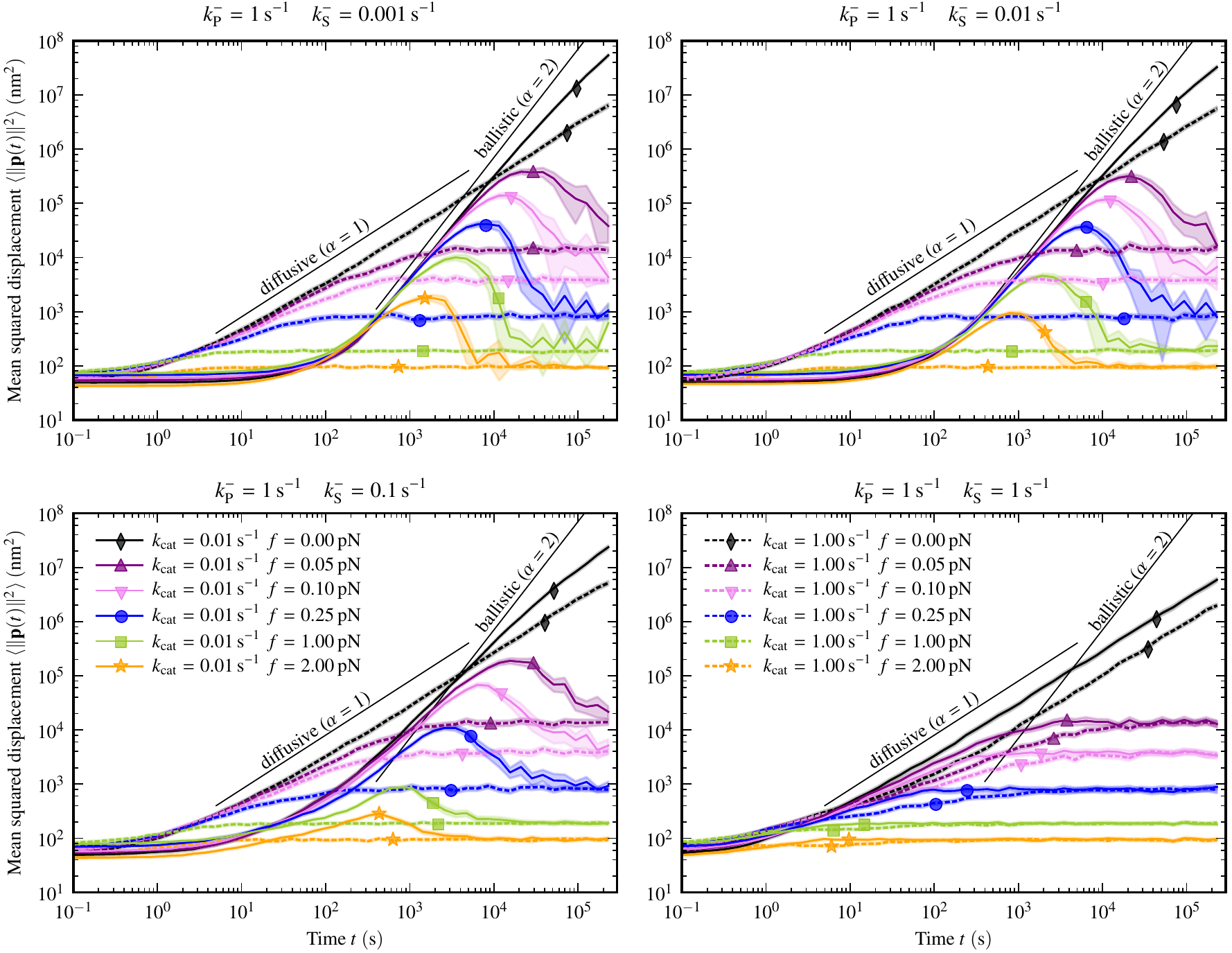}
    \caption{(color online) Simulation estimates ($n=1000$)
    of the mean squared displacement of the walkers as $\koffS$ is varied.
    Shading shows
    95\% confidence intervals for the mean.  Each subplot shows the same
    12 walker configurations, varying only the values of $\koffS$.  Fiducial
    lines for diffusion and ballistic motion are shown in the same position
    on each subplot for reference.
    These data can be compared with
    Figure~\ref{fig:forces-MSDLogLog} which shows the case $\koffS=0$.
    The value of $\koffS$ determines the rate of detachment without enzymatic
    conversion of the site to a product.  As long as $\koffS+\kcat < \koffP$,
    there remains a residence time bias, and
    our results show that walkers with $\kcat=0.01\pers$ move superdiffusivly
    for $\koffS<\koffP=1\pers$.
    This shows that the qualitative behavior of the walkers is unchanged for
    small variations in $\koffS$, and the choice of $\koffS=0$ in the model is appropriate
    as small values of $\koffS$ do not significantly affect the walker motion.
    However, when $\koffP=\koffS=1\pers$ the superdiffusive
    motion is eliminated, as there is no longer an effective residence time
    bias between visited and unvisited sites.\label{fig:16}}
\end{figure*}

We show the results of varying $\koffS$ in Fig.~\ref{fig:16}.  In other results
we have assumed that $\koffS=0$, which is reasonable as this rate is likely to
be much slower than $\koffP$ or $\kcat$ for any practical enzymatic implementation
of a multivalent random walker.  Figure~\ref{fig:16} shows that indeed the
superdiffusive behavior is robust to changes in $\koffS$, as long as it remains significantly
slower than $\koffP$ and $\kcat$.  However, setting $\koffS=\koffP$ eliminates
any superdiffusive effect, as there is no longer a residence time bias between
substrates and products, and the motion of the walker near the boundary is no longer
biased in the direction of the local substrate concentration gradient.

\subsection{Effect of forces on dissociation reactions}
In describing our model we show how forces affect the bimolecular
association rates through Eq.~\ref{eq:rB}.
However, applying a load force to walkers should also affect
the kinetics of the unimolecular dissociation events.  From
a high-level view, the kinetics of unimolecular reactions depend on a
molecule having enough internal energy to surmount some reaction
energy barrier $U_0$.  Thus, rate laws follow the Arrhenius formula,
$ k(T)\propto \exp{\left(-U_0/k_B T\right)}$.  The effect of a force
$f$ applied to the molecule is a mean change in energy of
$\Delta U_f$, and the rate is modified to

\begin{equation}
\label{eq:arrhenius}
k(T)= \nu \exp{\left((\Delta U_f-U_0)/k_B T\right)}.
\end{equation}

The value of the constant $\nu$ and the relationship of $\Delta U_f$
with force $f$ depend on the specific internal chemistry of the leg
tethers, enzymes, and substrates~\cite{Evans:1999, Willemsen:2000, Lee:2007},
the details of which are beyond the scope of our coarse-grained walker model.
We surmise that the effect of small forces
is a slight increase in $\koffS$ and $\koffP$, although this change would
not be uniform over all legs, as those attached to sites further in the $+\uvecx$
direction will oppose more of the load force on average than other sites.
Based on Fig.~\ref{fig:16}, small increases in
$\koffS$ do not qualitatively change the motive properties of the
walker with regard to MSD, except when the forces are large enough
that $\koffS+\kcat\geq\koffP$, which eliminates the residence time bias and
all superdiffusive motion.  Additionally, small increases in $\koffP$
actually lead to an increase in the MSD, as they increase the residence time
bias.  Quantitative analysis of variation in $\koffP$
is beyond the scope of this paper, but is available along with other supplementary
results for the MVRW model~\cite{Olah:2012:dissertation}.
Overall, these
results show that even though the present formulation of the MVRW model does not
describe the effect of force on dissociation rates,
we expect an extension of the model including these rates
to predict similar superdiffusive behaviors,
as long as the forces and corresponding rate changes are small.

\section{Conclusion}

The multivalent random walker model describes walker systems that can be designed to
act as translational molecular motors,  
without the need for complex intra-molecular conformational switching or gaiting.
Unlike models of natural molecular motors such as Kinesin I and Myosin V
~\cite{Astumian:2010,Lipowsky:2007,Linke:2009},
we do not assume any coordination between the legs,
chemically or mechanically.  Instead, we assume that the legs act independently.
The legs are passively constrained by their tethering
to a common body, but the chemical state of one leg cannot be communicated to the other
legs.  We show that such a simple walker design is able to exploit a residence time
bias in the enzymatic kinetics of the substrate and product sites it moves over
to generate a directional bias.
In addition, because the substrate sites take the dual role of
the chemical fuel source and the track binding sites, there is no need to couple
together in each foot a separate track binding and fuel binding site
as employed by kinesin I and other natural motors~\cite{Vale:2000}.
Hence, these difficult-to-engineer features that are found in natural
molecular motors are not strictly necessary for MVRW-like walkers
to transduce the chemical free energy of substrate catalysis into physical work,
and many enzyme/substrate
systems could provide the effective kinetics necessary for a multivalent random
walker to act as a molecular motor.
For this reason we avoid explicitly focusing on a
particular enzyme/substrate system in the MVRW model and instead
we explain how the interplay between the various kinetic rates
controls the ability of a multivalent random walker to act as a molecular motor,
transforming chemical free energy into directed motion and performing
physical work as it moves in opposition to a load force.

\begin{acknowledgments}
The authors would like to thank Milan N. Stojanovic, Cristopher Moore,
Lance R. Williams, Thomas P. Hayes, and Paul L. Krapivsky for helpful discussions
and advice regarding the development of our model and simulation software,
and the analysis and interpretation of results.
This material is based upon work supported by the National Science Foundation
under grants 0533065, 0829896, and 1028238.
\end{acknowledgments}

\end{document}